\documentclass[
reprint,
superscriptaddress,
nofootinbib,
amsmath,
amssymb,
aps,
prd,
floatfix
]{revtex4-2}
\usepackage[utf8]{inputenc}
\usepackage[T1]{fontenc}
\usepackage{slashed}

\newcommand{\edth}{\textnormal{\dh}}

\usepackage{mathrsfs}
\usepackage{fullpage}
\usepackage{enumitem}
\usepackage{mathalfa,amssymb,amsthm}
\usepackage{amsmath}
\usepackage{color}
\usepackage{amsmath}
\usepackage{graphicx}
\usepackage{dcolumn}
\usepackage{xcolor,subfigure,float}
\usepackage{bm}
\usepackage{amsmath}
\usepackage{cases,mathtools}
\usepackage{makecell}
\usepackage{verbatim} 
\usepackage{hyperref} 
\usepackage{cleveref}

\newcommand{\D}{\mathcal{D}}
\newcommand{\pd}{\partial}
\newcommand{\dd}{\mathrm{d}}
\newcommand{\ee}{\mathrm{e}}
\newcommand{\ii}{\mathrm{i}}
\newcommand{\Wj}[6]{\left(\begin{array}{ccc}#1&#2&#3\\#4&#5&#6\end{array}\right)}

\begin{document}
\title{Numerical computation of electromagnetically sourced nonlinear tails}
\author{Zhen-Tao He}
\email{hezhentao22@mails.ucas.ac.cn}
\affiliation{School of Physical Sciences, University of Chinese Academy of Sciences, Beijing 100049, China}

\author{Jia Du}
\affiliation{School of Physical Sciences, University of Chinese Academy of Sciences, Beijing 100049, China}

\author{Jiageng Jiao}
\email{jiaojiageng@ucas.ac.cn}
\thanks{Corresponding author}
\affiliation{International Centre for Theoretical Physics Asia-Pacific, University of Chinese Academy of Sciences, 100190 Beijing, China}

\author{Caiying Shao}
\email{shaocaiying@ucas.ac.cn}
\thanks{Corresponding author}
\affiliation{School of Physical Sciences, University of Chinese Academy of Sciences, Beijing 100049, China}

\author{Junxi Shi}
\affiliation{International Centre for Theoretical Physics Asia-Pacific, University of Chinese Academy of Sciences, 100190 Beijing, China}

\author{Yu Tian}
\email{ytian@ucas.ac.cn}
\thanks{Corresponding author}
\affiliation{School of Physical Sciences, University of Chinese Academy of Sciences, Beijing 100049, China}
\affiliation{\textit{Institute of Theoretical Physics, Chinese Academy of Sciences, Beijing 100190, China}}

\author{Hongbao Zhang}
\email{hongbaozhang@bnu.edu.cn}
\thanks{Corresponding author}
\affiliation{School of Physics and Astronomy, Beijing Normal University, Beijing 100875, China}
\affiliation{Key Laboratory of Multiscale Spin Physics, Ministry of Education, Beijing Normal University, Beijing 100875, China}
\date{\today}
\begin{abstract}
Amazingly, recent studies indicate that nonlinear effects are of great significance for modeling black hole ringdown.
Transient electromagnetic events in the astrophysical environment are typically high energetic, potentially responsible for some nonlinearities in ringdown.
Motivated by the desire to understand these nonlinearities, we solve the inhomogeneous Bardeen-Press-Teukolsky equation numerically and find second-order gravitational tails induced by an electromagnetic source.
Our results suggest that the second-order tails of curvature perturbations with multipole numbers $l\geq4$ decay as $ t^{-2l-2}$ at fixed spatial position and $u^{-l-3}$ in retarded-time $u$ at null infinity, slower than their linear counterparts, which can play a role in multi-messenger observations.
\end{abstract}
\maketitle

\section{Introduction}
The birth of gravitational waves (GWs) astronomy heralds a new era of gravitational physics, deepening our understanding of fundamental interactions in the strong-field regime \cite{Berti2015,Berti2018,Berti2018a,Franciolini2019}.
GWs from merging binary black holes (BHs), observed by the ground-based LIGO-Virgo-KAGRA detectors (also the forthcoming space-based missions such as LISA \cite{AmaroSeoane2017},  Taiji \cite{Hu2017}, and TianQin \cite{Luo2016}), allow us to test the validity of general relativity \cite{Abbott2016,Abbott2016a,Abbott2016b,Abbott2021a,Abbott2023a,Abac2024}.
The BH spectroscopy \cite{Berti2006,Berti2016,Berti2025} is such a program concerned with the ringdown stage, where the remnant relaxes to a stationary BH. 

In the BH ringdown, perturbative radiation fields relax in a set of characteristic quasi-normal modes (QNMs) and then are dominated by inverse power-law tails, which have been extensively explored in the linear BH perturbation theory (see e.g., \cite{Price1972,Price1972a,Leaver1986,Gundlach1994a,Price1994,Krivan1997,Hod1999,Barack1999,Barack1999a,Hod2000,Hod2000a,Burko2003,Zenginoglu2008a,Zenginoglu2010,Racz2011,Harms2013,Csukas2019,Csukas2021,Rosato2025}). 
Recently, however, increasing studies reveal rich phenomena arising from nonlinearities in BH ringdown, such as quadratic QNMs \cite{Mitman2023,Cheung2023,Lagos2023,Baibhav2023,Bucciotti2023,Khera2023,Perrone2024,RedondoYuste2024,Yi2024,Ma2024,Zhu2024,Qiu2024,Bucciotti2024,May2024,Bucciotti2024a,Pan2024,Cheung2024,Bourg2025,Giesler2025,Khera2025,Kehagias2025b,Lagos2025,Aly2025,Aly2024,Kehagias2025c,Perrone2025} and nonlinear power-law tails  {\cite{Cardoso2024,Ling2025,Kehagias2025,Kehagias2025a,DeAmicis2024,Ma2025,Alvares2025,Ianniccari2025}}.
Surprisingly, not only can the quadratic QNM’s amplitude be even larger than that of their linear counterpart \cite{Mitman2023,Cheung2023}, but the nonlinear tails can decay more slowly than the linear Price's law \cite{Cardoso2024,Ling2025,Kehagias2025,Kehagias2025a,DeAmicis2024,Ma2025}.
These results indicate that the linear BH perturbation theory alone is insufficient to model BH ringdown accurately, necessitating the nonlinear corrections. 

Transient electromagnetic (EM) events in the astrophysical environment are typically high energetic, e.g., short gamma-ray bursts (GRBs) with energy released ranging from $10^{50}$ to $10^{52}$ \cite{Wu2012,Berger2014}, supernovae events, or long GRBs releasing energy of around $10^{54}$ ergs \cite{Jakobsson2006,Srinivasaragavan2023}.
Moreover, if a neutron star has magnetic energy up to $10^{49}$ ergs and a mass of about one solar mass $M_\odot$, then its EM effect could surpass its second-order gravitational effect in extreme mass ratio inspiral systems with a supermassive BH $M\gtrsim 10^{6}M_\odot$ \cite{Aly2025}.
Such high energy renders EM sources potentially responsible for some nonlinearities in ringdown \cite{Aly2025,Aly2024}.
Hence, multi-messenger astronomy \cite{Abbott2017,Neronov2019}, which integrates GW detection and EM observation, could help the interpretation of GW signals.
For example, by analyzing the QNMs of GWs generated from transient high-energetic EM waves, one can detect solitary black holes, whose number and distribution in the Milky Way encode essential information about BH formation and the existence of primordial BHs \cite{Jana2024}.

Then a natural question arises: \textit{``Can we further the understanding or observation of the electromagnetically sourced  nonlinearities in BH ringdown?''}
Inspiringly, the analytic calculation for an ideal dipole radially free falling into a Schwarzschild BH in \cite{Aly2025} suggests a sign of polynomial tails in gravitational perturbations sourced by electromagnetic perturbations.
However, conclusive evidence still needs to be found.

To this end, we investigate second-order gravitational perturbations sourced by first-order electromagnetic perturbations under the Schwarzschild spacetime.
We find that electromagnetically sourced nonlinear tails do exist by numerically solving the Bardeen-Press-Teukolsky (BPT) equations in the Newman-Penrose (NP) formalism which can be easily extended to Kerr spacetime \cite{Bardeen1973,Teukolsky1973}.

The paper is organized as follows.
In Section \ref{sec2}, we describe our methodology to study the electromagnetically sourced nonlinear tails, including hyperboloidal foliations, reconstruction of the Maxwell scalars, and calculations of quadratic source terms.
Our numerical scheme, presented in Section \ref{sec3}, is comprised of spatial discretization, analytical mesh refinement, a first-order time reduction, and time-symmetric integration.
Detailed numerical results are shown in Section \ref{sec4}, including discussions of initial data and mode coupling, the power law of second-order tails and its astrophysical implications, and numerical checks.
Finally, in Section \ref{sec5}, we sum up our concluding remarks.

In this paper, we use the geometric units $c=8\pi G=1$ and the conventions of Chandrasekhar \cite{Chandrasekhar1998}.
For example, the metric signature is $(+,-,-,-)$ and the complex conjugate of a quantity $f$ is denoted as $\bar{f}$.
However, we use Greek letters as spacetime indices instead.
Furthermore, the original NP spin coefficient $\pi$ is denoted by a variant pi ``$\varpi$'' to avoid confusion, and a variant epsilon ``$\varepsilon$'' denotes a perturbative parameter distinguished from the original NP spin coefficient ``$\epsilon$.''
Finally, an $n$th order quantity in a perturbative expansion is denoted with a trailing superscript ${}^{(n)}$, e.g., $\phi=\phi^{(0)}+\varepsilon\phi^{(1)}+\varepsilon^2\phi^{(2)}+\cdots$.

All codes used to create data corresponding to the findings in this manuscript are openly available \cite{mydata}.

\section{Methodology} \label{sec2}
In this Section, we present the inhomogeneous BPT equation in horizon-penetrating hyperboloidally compactified coordinates, along with the reconstruction of Maxwell scalars via the Maxwell equations, which are required for the calculation of the quadratic source term.
\subsection{Newman-Penrose formalism and hyperboloidal foliations}
We consider a Schwarzschild BH with mass $M$ that is perturbed by a sourceless electromagnetic field in the form of the Maxwell scalar in the NP formalism $\phi_2=\varepsilon\phi^{(1)}_2+\mathcal{O}(\varepsilon^2)$.
Because the energy-momentum tensor of the Maxwell field $T_{\mu\nu}^{\text{EM}}$ is a quadratic form of the Maxwell scalars $\phi$, the gravitational perturbation is nonlinearly sourced at the leading order $\Psi_4=\varepsilon^2\Psi_4^{(2)} + \mathcal{O}(\varepsilon^3)$.
Thus, the system of equations under consideration is
\begin{equation} \label{teukm1}
    {}_{-1}\mathcal{T}\phi^{(1)}_2={}_{-1}\mathcal{S}=0,
\end{equation}
\begin{equation} \label{teukm2}
    {}_{-2}\mathcal{T}\Psi_4^{(2)}={}_{-2}\mathcal{S}(\phi^{(1)}_2;\phi^{(1)}_2),
\end{equation}
where ${}_{s}\mathcal{T}$ is the Teukolsky operator for the NP scalar with a spin-weight $s$.
The source term ${}_{-2}\mathcal{S}$ reads \cite{Loutrel2021}
\begin{equation}
    {}_{-2}\mathcal{S} 
    =d_{4}^{(0)}\mathcal{R} _{d}^{(2)}+d_{3}^{(0)}\mathcal{R} _{h}^{(2)},
\end{equation}
where the Ricci terms $\mathcal{R} _{d}^{(2)}$ and $\mathcal{R} _{h}^{(2)}$ are
\begin{equation} \label{Rd}
\begin{aligned}
    \mathcal{R} _{d}^{(2)}
    =&
    (\bar{\delta}+2\alpha - 2\bar{\tau})^{(0)}\Phi_{21}^{(2)}
    \\
    &-(\Delta +\bar{\mu}+2\gamma-2\bar{\gamma})^{(0)}\Phi _{20}^{(2)},
    \end{aligned}
\end{equation}

\begin{equation}
\begin{aligned}
\mathcal{R} _{h}^{(2)}
    =&
    (\Delta +2\bar{\mu} +2\gamma)^{(0)}\Phi _{21}^{(2)}
    \\
    &-(\bar{\delta}+2\alpha +2\bar{\beta}-\bar{\tau})^{(0)}\Phi _{22}^{(2)},
\end{aligned}
\end{equation}
and two derivatives $d_3$ and $d_4$ are defined as 
\begin{align}
	d_3&\equiv \bar{\delta}+3\alpha +\bar{\beta}
	+4\varpi -\bar{\tau},\\
	d_4&\equiv \Delta +4\mu +\bar{\mu}
	+3\gamma -\bar{\gamma}.
\end{align}
As $T_{\mu\nu}^{\text{EM}}$ is traceless, one can obtain the Ricci scalars $\Phi_{mn}^{(2)} = -\phi_m^{(1)}\bar{\phi}_n^{(1)}$ from the Einstein equation $R_{\mu\nu} = -T_{\mu\nu} $.

To avoid complicated boundary condition problems \cite{Dafermos2004,Zenginoglu2008} and to extract astrophysically relevant results at future null infinity, we use horizon-penetrating hyperboloidally compactified coordinates $\{T,R,\theta,\varphi\}$ in the minimal gauge \cite{Macedo2018,Macedo2019} and a rotated Kinnersley tetrad [see \cref{l,n,m}], which is regular on the horizon \cite{Ripley2021}.
Specifically, the transformation from the Schwarzschild coordinates $\{t,r,\theta,\varphi\}$ to the hyperboloidal coordinates $\{T,R,\theta,\varphi\}$ is
\begin{equation}
\begin{aligned}
    & T = t - r + 2M\ln\frac{r-2M}{2M} -4M\ln\frac{r}{2M},
    \\
    & R = \frac{L^2}{r},
\end{aligned}
\end{equation}
with $L$ a constant parameter which, as well as $M$, is set to 1 in this work.
Future null infinity $\mathcal{I}^+$ is located at $R=0$, and the future event horizon $\mathcal{H}^+$ is located at $R_\text{H}=L^2/2M$.
Note that the hyperboloidal time coordinate $T$ approaches a retarded time $u$ when $R\to0$ and approaches an advanced time $v$ when $R\to R_\text{H}$.

In our coordinates and tetrad, the BPT equation reads
\begin{widetext}
\begin{equation} \label{mastereq}
    \begin{aligned}
    \left(
	C_{TT} \partial _{T}^{2}
	+ C_{TR} \partial_T\partial_R
	+C_{RR} \partial _{R}^{2}
	+{}_{s}C_T  \partial _T
	+{}_{s}C_R \partial _R
	+{}_{s}C  
	-{}_s\slashed{\Delta}
	\right)
	{}_{s}\hat{\psi}
	= {}_s\hat{\mathcal{S}},
    \end{aligned}
\end{equation}
\end{widetext}
where the equation coefficients read
\begin{align}
    C_{TT} &= 16M^2\left( 1+\frac{2MR}{L^2} \right),
    \\
    C_{TR} &= -2\left( L^2-\frac{8M^2R^2}{L^2}\right),
    \\
    C_{RR} &= -\left( L^2-2MR \right) \frac{R^2}{L^2},
\end{align}
\begin{align}
    {}_{s}C_T &= 4M\left[ -s+(2+s)\frac{2MR}{L^2} \right],
    \label{cT}
    \\
    {}_{s}C_R &= 2R\left[ -(1+s)+(s+3)\frac{MR}{L^2} \right],
    \\
    {}_{s}C   &= 2 (1+s)\frac{MR}{L^2},
\end{align}
and ${}_s\slashed{\Delta}$ is the spin-weight $s$ Laplace-Beltrami operator on the unit two-sphere.
The rescaled master function ${}_{s}\hat{\psi}$ and rescaled source term ${}_{s}\hat{\mathcal{S}}$ are related to the NP scalars (see Table \ref{masterfun}). 

\begin{table}\label{masterfun}
    \centering
    \begin{tabular}{c|c|c}
    \hline
    $s$ & ${}_{s}\hat{\psi}$ 
    & ${}_{s}\hat{\mathcal{S}}$ 
    \\
    \hline
        -2 & $\Psi_{4}/R$
        & $(2L^4/R^3){}_{-2}\mathcal{S}$
        \\
        -1 & $\phi_{2}/R$
        & $(2L^4/R^3){}_{-1}\mathcal{S}$
        \\
        +1 & $\phi_{0}(\Psi_2/M)^{-2/3}/R$
        & $(2L^4/R^3)(\Psi_2/M)^{-2/3}
        {}_{+1}\mathcal{S}$
        \\
        +2 & $\Psi_{0}(\Psi_2/M)^{-4/3}/R$
        & $(2L^4/R^3)(\Psi_2/M)^{-4/3}
        {}_{+2}\mathcal{S}$
        \\
        \hline
    \end{tabular}
    \caption{Relation between the rescaled scalars ${}_{s}\hat{\psi}, {}_{s}\hat{\mathcal{S}}$ and the NP scalars.}
\end{table}
Due to spherical symmetry of the Schwarzschild background, we expand ${}_{s}\hat{\psi}$, as well as ${}_{s}\hat{\mathcal{S}}$, in terms of the spin-weight spherical harmonics ${}_{s}Y_{lm}(\theta,\varphi)$,
the eigenfunction with an eigenvalue $-(l-s)(l+s+1)$ of ${}_s\slashed{\Delta}$,
\begin{equation}
    {}_{s}\hat{\psi} (T,R,\theta,\varphi) = \sum_{l,m}{}_{s}\hat{\psi}^{[lm]}(T,R) {}_{s}Y_{lm}(\theta,\varphi).
\end{equation}
Substituting this expansion into the master equation \eqref{mastereq} gives a series of decoupled (1+1)-dimensional partial differential equations (PDEs) for each mode ${}_{s}\hat{\psi}^{[lm]}$, where the characteristic speeds read
\begin{equation}
\begin{aligned}
    c_{\pm} &= \frac{ C_{TR}\mp\sqrt{C_{TR}^2-4C_{RR}C_{TT}} } {2C_{TT}}
    \\
    &=
    \frac
    {8 M^2 R^2-L^4 \mp L^4}
    {16 M^2 \left(L^2+2 M R\right)}.
\end{aligned}
\end{equation}
Note that the ingoing one $c_-$ vanishes at the future null infinity $(R=0)$, while the outgoing one $c_+$ vanishes at the future event horizon $(R=L^2/2M)$, for which the physical quasi-normal boundary conditions are inherently satisfied.

\subsection{Reconstruction of the Maxwell scalars}
Before calculating the source term ${}_s\hat{\mathcal{S}}$, we need to solve, via the Maxwell equations, other Maxwell scalars $\phi_1^{(1)}$ and $\phi_0^{(1)}$ (also their derivatives) consistent with the solution $\phi_2^{(1)}$ to eq. \eqref{teukm1}.
In this and next subsection, we will drop the spin coefficients that vanish in our tetrad (see \ref{sec_tetrad}).

The Maxwell scalar $\phi^{(1)}_1$ is reconstructed from the following Maxwell equation

\begin{equation} \label{recon1}
    \bar{\delta} \phi^{(1)}_1
    =
    \frac{R}{\sqrt{2}L^2} \edth^\prime \phi_1^{(1)}
    =
    (D -\rho + 2\epsilon )^{(0)} \phi^{(1)}_2.
\end{equation}
Note that the $\edth^\prime$ operator lowers the spin weight by 1. \footnote{For example, $\edth^\prime{}_{s}Y_{lm}=-\sqrt{(l+s)(l-s+1)}{}_{s-1}Y_{lm}$.
See \cite{Goldberg1967} for detailed discussion on the $\edth$ operator.}
That is, the $(l,m)$ mode of $\phi_1^{(1)}$ can only be excited by the same $(l,m)$ mode of $\phi_2^{(1)}$.
Thus, if we expand $\phi_1^{(1)}$ in terms of $Y_{lm}$
\begin{equation}
    \phi_1^{(1)} = \sum_{l,m}\phi_1^{[lm]}(T,R) Y_{lm},
\end{equation}
and substitute this into eq. \eqref{recon1}, then we obtain
\begin{equation}
\begin{aligned}
    \phi_1^{[lm]}
    &=
    -\sqrt{ \frac{2}{l(l+1)}}
    \left[
    \frac{R}{L^2}\left(
    \hat{l}^T \pd_T + \hat{l}^R \pd_R
    \right)
    +\frac{1}{2}
    \right]
    \phi_2^{[lm]}
    \\
    &=
    -\sqrt{ \frac{2}{l(l+1)}}
    \frac{R^2}{L^2}
    \left[
    \left(
    \hat{l}^T \pd_T +
    \hat{l}^R \pd_R
    \right)
    + M
    \right]
    {}_{-1}\hat{\psi}^{[lm]},
\end{aligned}    
\end{equation}
where $\hat{l}^T=4M^2$ and $\hat{l}^R=-(L^2-2MR)/2$.
Note that the fall-off behavior $\phi_1 \sim R^2$ is important for computing the rescaled source $\hat{\mathcal{S}}$ numerically.

The directional derivative $n^\mu\nabla_\mu \phi_1^{(1)}=\Delta \phi_1^{(1)}$ is reconstructed from
\begin{equation}
    (\Delta+2\mu)^{(0)} \phi_1^{(1)}
    = 
    (\delta+2\beta)^{(0)} \phi_2^{(1)}
    =
    \frac{R}{\sqrt{2}L^2} \edth \phi_2^{(1)},
\end{equation}
i.e.,
\begin{equation}
\begin{aligned}
    \Delta \phi_1^{[lm]}
    & = 
    \frac{R}{L^2}
    \left[
    2\phi_1^{[lm]}
    +
    \sqrt{\frac{l(l+1)}{2}} 
    \phi_{2}^{[lm]}
    \right]
    \\
    & = 
    \frac{R^2}{L^2}
    \left[
    2R
    \left(
    \frac{\phi_1^{[lm]}}{R^2}
    \right)
    +
    \sqrt{\frac{l(l+1)}{2}} 
    {}_{-1}\hat{\psi}^{[lm]}
    \right],
\end{aligned}
\end{equation}
where we use the $\edth$ operator's property of raising the spin weight by 1, i.e.,
$
    \edth{}_{s}Y_{lm}=\sqrt{(l-s)(l+s+1)}\,{}_{s+1}Y_{lm}.
$
The fall-off behavior of $\Delta\phi_1$ is also $\sim R^2$.

The reconstruction equation of $\phi_0$ reads
\begin{equation} \label{recon0}
   (\D - 2\rho)^{(0)}\phi_1^{(1)} = (\bar{\delta}-2\alpha)^{(0)}\phi_0^{(1)}= \frac{R}{\sqrt{2}L^2} \edth^\prime \phi_0^{(1)}.
\end{equation}
 {
Similarly, if we expand $\phi_0^{(1)}$ in terms of ${}_{1}Y_{lm}$
\begin{equation}
    \phi_0^{(1)} = \sum_{l,m}\phi_0^{[lm]}(T,R)\, {}_{1}Y_{lm}(\theta,\varphi)
    ,
\end{equation}
}
then eq. (\ref{recon0}) reads
\begin{equation}
\begin{aligned}
    \phi_0^{[lm]}
    & =
    -\sqrt{ \frac{2}{l(l+1)}}
    \frac{1}{L^2}
    \left[
    R\left(
    \hat{l}^T \pd_T +
    \hat{l}^R \pd_R
    \right)
    -2\hat{l}^R
    \right]
    \phi_1^{[lm]}
    \\
    & =
    -\sqrt{ \frac{2}{l(l+1)}}
    \frac{R^3}{L^2}
    \left(
    \hat{l}^T \pd_T +
    \hat{l}^R \pd_R
    \right)
    \left(
    \frac
    { \phi_1^{[lm]} }
    {R^2}
    \right).
\end{aligned}
\end{equation}

Finally, the reconstruction equation of $\Delta\phi_0$ reads
\begin{equation}
    \delta\phi_{1}^{(1)}=(\Delta+\mu)^{(0)}\phi_{0}^{(1)},
\end{equation}
i.e., 
\begin{equation} \label{nphi0}
\begin{aligned}
    \Delta \phi_0^{[lm]} 
    &= 
    \frac{R}{L^2}
    \left[
      \phi_0^{[lm]}
    + 
    \sqrt{\frac{l(l+1)}{2}}
    \phi_1^{[lm]}
    \right]
    \\
    &=
    \frac{R^3}{L^2}
    \left[
     R
     \left(
     \frac{\phi_0^{[lm]}}
     {R^3}
     \right)
    + 
    \sqrt{\frac{l(l+1)}{2}}
    \left(
    \frac{ \phi_1^{[lm]} }
    {{R^2}}
    \right)
    \right].
\end{aligned}
\end{equation}

Note that the second-order derivative $\Delta^2\phi_0$ is also required to compute ${}_{-2}\mathcal{S}$, which can be obtained from eq. \eqref{nphi0} 
\begin{equation}
    \Delta^{2}\phi_{0}^{[lm]}=\frac{R^3}{L^{2}}
    \left[
    2R\frac{\Delta\phi_{0}^{[lm]}}{R^{3}}+\sqrt{\frac{l(l+1)}{2}}\frac{\Delta\phi_{1}^{[lm]}}{R^{2}}
    \right].
\end{equation}
The falloff behavior of $\phi_0$, $\Delta\phi_0$ and $\Delta^2\phi_0$ are all $\sim R^3$.
\subsection{Quadratic source}

As mentioned above, the rescaled source ${}_{-2}\hat{\mathcal{S}}$ in \eqref{mastereq} is expanded in terms of ${}_{-2}Y_{lm}$
\begin{equation}
    {}_{-2}\hat{\mathcal{S}}
    =
    \sum_{l_3, m_3}
    {}_{-2}\hat{\mathcal{S}}^{[{l_3 m_3}]}
    {}_{-2}Y_{l_3 m_3}.
\end{equation}
For convenience, we divide it into three parts 
\begin{equation} \label{Sres}
    {}_{-2}\hat{\mathcal{S}}^{[{l_3 m_3}]}
    =
    \hat{\mathcal{S}}_1^{[{l_3 m_3}]}
    -
    \hat{\mathcal{S}}_2^{[{l_3 m_3}]}
    -
\hat{\mathcal{S}}_3^{[{l_3 m_3}]},
\end{equation}
where\footnote{A property ${}_{s}\bar{Y}_{lm}=(-1)^{m+s}{}_{-s}{Y}_{l,-m}$ is utilized in the following derivation.}
\begin{widetext} 
\begin{equation}\label{S1}
\begin{aligned}
\hat{\mathcal{S}}_1^{[{l_3 m_3}]}
=
&
\left\{
\frac{2 L^4}{R^3}
\left[
    d_4^{(0)}
    (\bar{\delta}+2\alpha )^{(0)}
    +
    d_3^{(0)}
    (\Delta +2\bar{\mu} )^{(0)}
\right]\Phi _{21} ^{(2)}
\right\}
^{[l_3m_3]}
= 
\sqrt{2}L^2\sum_{l_1m_1}\sum_{l_2m_2}
(-1)^{m_2}
\\
&\times
\left[
\sqrt{(l_1 -1)(l_1+2)}
{}_{-2\,0\,2\,}G_{l_1}^{\,m_1}{}_{l_2}^{-m_2}{}_{l_3}^{-m_3}
+
\sqrt{l_2(l_2+1)}
{}_{-1\,-1\,2\,}G_{l_1}^{\,m_1}{}_{l_2}^{-m_2}{}_{l_3}^{-m_3}
\right]
\\
&\times
\left[
2\Delta\phi_2^{[l_1m_1]}
    \left(
    \frac
    { \bar{\phi}_1^{[l_2m_2]} }
    {R^2}
    \right)
+
2 R {}_{-1}\psi^{[l_1m_1]}
    \left(
    \frac
    { \Delta\bar{\phi}_1^{[l_2m_2]} }
    {R^2}
    \right)
+
{}_{-1}\psi^{[l_1m_1]}
    \left(
    \frac
    { \bar{\phi}_1^{[l_2m_2]} }
    {R^2}
    \right)
[n^{R}+(4\mu+3\bar{\mu})R]
\right],
\end{aligned}
\end{equation}

\begin{equation}\label{S2}
\begin{aligned}
\hat{\mathcal{S}}_2^{[{l_3 m_3}]}
=
    \left\{ \frac{2L^4}{R^3}d_{4}^{(0)}\left( \Delta +\bar{\mu} \right) ^{(0)}\Phi _{20} ^{(2)} \right\} ^{[l_3m_3]}
    =
    &2L^4\sum_{l_1m_1}{\sum_{l_2m_2}{\left( -1 \right) ^{m_2}}}\,\,_{-1\,-1\,2\,}
    {G_{l_1}^{\,m_1}}{}_{l_2}^{-m_2}{}_{l_3}^{-m_3}
\\
&
\times
\left[ 
	 \Delta ^2\phi _{2}^{[l_1m_1]}\frac{\bar{\phi}_{0}^{[l_2m_2]}}{R^3}+2\Delta \phi _{2}^{[l_1m_1]}\frac{\Delta \bar{\phi}_{0}^{[l_2m_2]}}{R^3}
	 +\phi _{2}^{[l_1m_1]}\frac{\Delta ^2\bar{\phi}_{0}^{[l_2m_2]}}{R^3} 
	\right.
	\\
	&+2\left( 2\mu +\bar{\mu} \right) \left( \Delta \phi _{2}^{[l_1m_1]}\frac{\bar{\phi}_{0}^{[l_2m_2]}}{R^3}+\phi _{2}^{[l_1m_1]}\frac{\Delta \bar{\phi}_{0}^{[l_2m_2]}}{R^3} \right)
	\\
	&\left.
	+\left[ \left( n^R\partial _R\bar{\mu} \right) +\left( 4\mu +\bar{\mu} \right) \bar{\mu} \right] \phi _{2}^{[l_1m_1]}\frac{ \bar{\phi}_{0}^{[l_2m_2]}}{R^3}
    \right] ,
\end{aligned}
\end{equation}
and
\begin{equation}\label{S3}
\begin{aligned}
&\hat{\mathcal{S}}_3^{[{l_3 m_3}]}
=
    \left\{ \frac{2L^4}{R^3}d_{3}^{(0)}(\bar{\delta}+2\alpha +2\bar{\beta})\Phi _{22}^{(2)} \right\} ^{[l_3m_3]}
    =R\sum_{l_1m_1}{\sum_{l_2m_2}{\left( -1 \right) ^{m_2}}}
\times
\left[ 
l_2( l_2+1 ) \,_{-1\,-1\,2\,}
{G_{l_1}^{\,m_1}}{}_{l_2}^{-m_2}{}_{l_3}^{-m_3}+
\right.
\\
&\left.
\sqrt{\left( l_1-1 \right) \left( l_1+2 \right)}\left( 2\sqrt{l_2\left( l_2+1 \right)}
{}_{-2\,0\,2\,}{G_{l_1}^{m_1}}{}_{l_2}^{-m_2}{}_{l_3}^{-m_3}
+\sqrt{\left( l_1-2 \right) \left( l_1+3 \right)}_{-3\,1\,2\,}{G_{l_1}^{m_1}}{}_{l_2}^{-m_2}{}_{l_3}^{-m_3} \right) \right]
{}_{-1}{\hat{\psi}^{[l_1m_1]}}
{}_{-1}
\bar{\hat{\psi}}^{[l_2m_2]}.
\end{aligned}
\end{equation}
Here, the mode coupling is determined by the corresponding Gaunt coefficients 
\begin{equation}
{}_{s_1s_2s_3\,}G_{\ell_1}^{\,m_1}{}_{\ell_2}^{m_2}{}_{\ell_3}^{m_3}=\int {}_{s_1}Y_{\ell_1m_1}\cdot{}_{s_2}Y_{\ell_2m_2}\cdot{}_{s_3}Y_{\ell_3m_3}\sin\theta\,\mathrm{d}\theta\mathrm{d}\varphi\,,
\end{equation}
which can be expressed via the Wigner-3j symbols as

\begin{equation}
    {}_{s_1\,s_2\,s_3\,}G_{\ell_1}^{\,m_1}{}_{\ell_2}^{m_2}{}_{\ell_3}^{m_3}=\sqrt{\frac{(2\,\ell_1+1)(2\,\ell_2+1)(2\,\ell_3+1)}{4\pi}}\\
    \times
    \Wj{\ell_1}{\ell_2}{\ell_3}{-s_1}{-s_2}{-s_3}\Wj{\ell_1}{\ell_2}{\ell_3}{m_1}{m_2}{m_3}\,.
\end{equation}
This completes our derivation of the inhomogeneous BPT equation with an electromagnetic source.
\end{widetext}


\section{Numerical scheme} \label{sec3}
In this Section, we describe our numerical scheme to solve the (1+1)-dimensional PDEs, i.e., eq. \eqref{mastereq} with ${}_s\slashed{\Delta}$ replaced by the eigenvalue $-(l-s)(l+s+1)$.

High-precision floating-point numbers are required to obtain accurate tail decay rates.
In our numerical simulations, we employ \textsf{DoubleFloat64} floating-point numbers provided by the Julia package \textsf{DoubleFloats.jl} \cite{Sarnoff2022}.

\subsection{Spatial discretization and analytical mesh refinement}
For spatial discretization, the Chebyshev pseudo-spectral method is used in the radial $R$ direction. 
However, growing gradients close to the future null infinity $\mathcal{I}^+$ occur in the late-time profile of the master function ${}_{s}\hat{\psi}^{[lm]}$ due to the differences between the decay rates at $\mathcal{I}^+$ and finite radii.
We use analytical mesh refinement (AnMR) to solve this problem.

Specifically, $N+1$ scaled and shifted Chebyshev-Gauss-Lobatto collocation points $\{R_i ^{\text{Cheb}}\}$ are used within $R\in[0,R_\text{H}]$ during the quasi-normal ringing phase,
\begin{equation}
    R_i ^{\text{Cheb}} = \frac{R_\text{H}}{2}\left(1+\cos\frac{i\pi}{N}\right),\, i=0,...N.
\end{equation}

While ${}_{s}\hat{\psi}^{[lm]}(T,R)$ enters the polynomial tail decay phase, its profile behaves like a function (see e.g., Fig. 9 in \cite{Macedo2014})
\begin{equation} \label{func}
    \psi(R) = \frac{a}{|a|+R}, \quad 
     {
    |a|}\ll1,
\end{equation}
whose $n$th derivative at $R=0$ is of order $|a|^{-n}$.
Hence, a new coordinate $\sigma$ is introduced instead 
\begin{equation} \label{AnMR}
    R(\sigma) = R_{\text{H}} \frac{\sinh[\kappa (\sigma+1)/2]}{\sinh\kappa},
\end{equation}
with a mesh-refinement parameter \cite{Meinel2008,Macedo2014}
\begin{equation} \label{AnMR_kappa}
    \kappa = \ln |a| = \ln\left| 
    \frac { {\psi} }  { {\psi}_{,R} }
    \right|_{R=0},
\end{equation}
in which the $n$th derivative of the function \eqref{func} at $R=0$ merely goes as $(\ln |a|)^n$.
We monitor behaviors of ${}_{s}\hat{\psi}^{[lm]}$ during evolutions and set the parameter $\kappa$ by substituting ${}_{s}\hat{\psi}^{[lm]}$ into eq. \eqref{AnMR_kappa} once ${}_{s}\hat{\psi}^{[lm]}$ evolves into a configuration similar with eq. \eqref{func}. \footnote{${}_{-1}\hat{\psi}^{[lm]}(T,R)$ enters the tail phase typically earlier than ${}_{-2}\hat{\psi}^{[lm]}(T,R)$.} 

\begin{figure}[htp]
    \centering
    \includegraphics[width=0.46\textwidth]{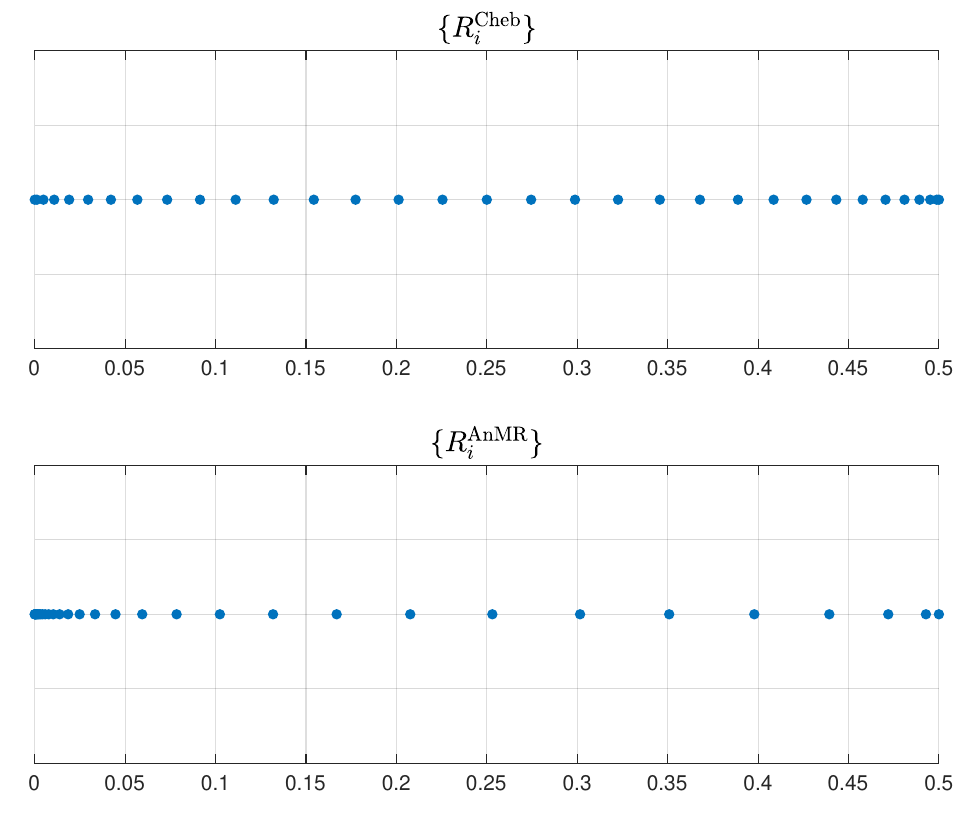}
    \caption{Comparison between the collocation points $\{R_i^{\text{Cheb}}\}$ and $\{R_i^{\text{AnMR}}\}$. 
    Here, we set $\kappa=6$ and $N=N^\prime=32$.}
    \label{grid}
\end{figure}

\begin{figure}[htp]
    \centering
    \includegraphics[width=0.46\textwidth]{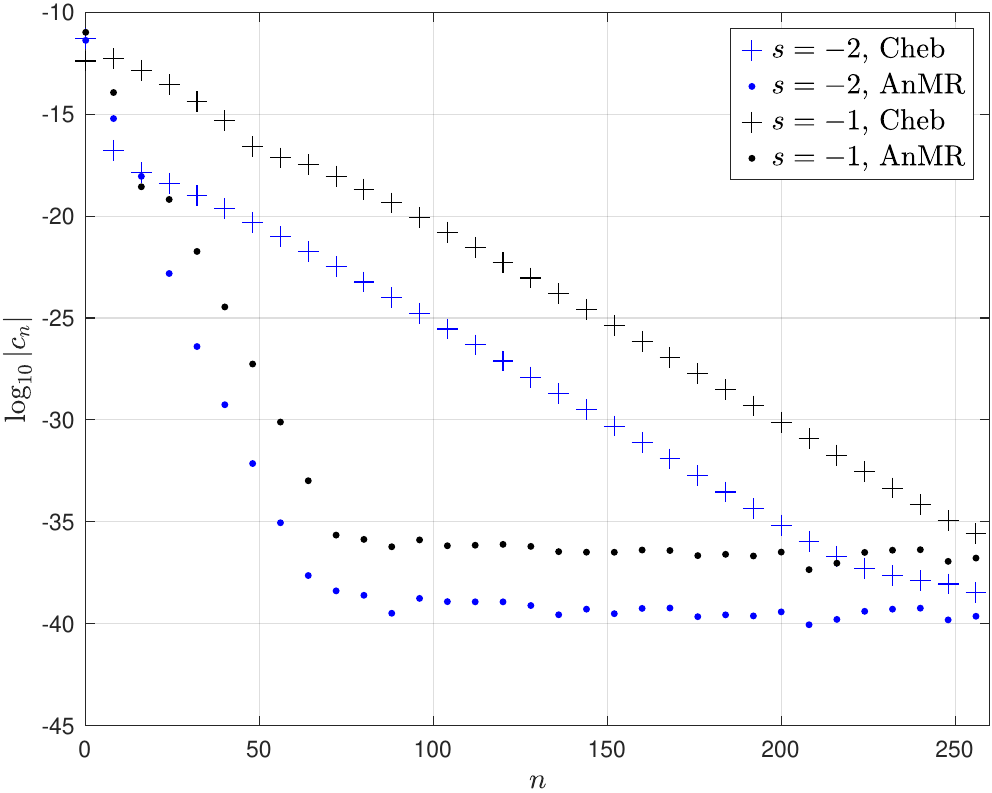}
    \caption{Spectral coefficients $c_n$ of $\{{}_{-1}\hat{\psi}^{[33]}(T,R_i)\}$ and $\{{}_{-2}\hat{\psi}^{[55]}(T,R_i)\}$ for both the grids $\{R_i^{\text{Cheb}}\}$ and $\{R_i^{\text{AnMR}}\}$ when a steep gradient of ${}_{-1}\hat{\psi}^{[33]}$ occurs at about $T=260M$.
    Here, we take $N=N^\prime=256$.
    }
    \label{spec_vs}
\end{figure}

Using the AnMR \eqref{AnMR}, we map $N^\prime+1$ standard Chebyshev-Gauss-Lobatto collocation points $\{\sigma_i^{\text{Cheb}}\}$ within $\sigma\in[-1,1]$ into the interval $R\in[0,R_{\text{H}}]$.
The AnMR collocation points $\{R_i^{\text{AnMR}}\}=\{R(\sigma_i^{\text{Cheb}})\}$ are more dense near $R = 0$ than $\{R_i^{\text{Cheb}}\}$, \footnote{A drawback of AnMR is that $\{R_i^{\text{AnMR}}\}$ are sparse near the other boundary $R=L^2/2M$, for which we do not employ AnMR from the beginning of evolutions.} as shown in Fig. \ref{grid}.
When transforming from $\{R_i^{\text{Cheb}}\}$ to $\{R_i^{\text{AnMR}}\}$, we interpolate the field ${}_{s}\hat{\psi}^{[lm]}$ in terms of Chebyshev polynomials of the first kind via its grid values $\{{}_{s}\hat{\psi}^{[lm]}(R_i^{\text{Cheb}})\}$ to obtain its values at the collocation points $\{R_i^{\text{AnMR}}\}$. \footnote{ {Altering $\{R_i^{\text{AnMR}}\}$ by readjusting $\kappa$ also involves such a spectral interpolation, which will significantly increase computing costs if one readjusts $\kappa$ at each time step. Thus, updating $\kappa$ before the spectral convergence is spoiled will suffice.}}
As shown in Fig. \ref{spec_vs}, spectral coefficients $c_n$ of the numerical solution $\{{}_{s}\hat{\psi}^{[lm]}(T,R_i^{\text{AnMR}})\}$ converge faster than those of $\{{}_{s}\hat{\psi}^{[lm]}(T,R_i^{\text{Cheb}})\}$, which reduces computing costs to get accurate tail behaviors.

\subsection{First-order time reduction}
With the above spatial discretization, the spatial derivative operator $\partial_R$ is replaced by a differentiation matrix $\bm{D}$ \cite{Trefethen2000}. 
Then, the original PDEs of fields ${}_{s}\hat{\psi}^{[lm]}(T,R)$ turn into a system of coupled second-order ordinary differential equations (ODEs) of the value of the fields at the $N^{(\prime)}+1$ collocation points $\{{}_{s}\hat{\psi}^{[lm]}(T,R_i)\}$, where $N^{(\prime)},R_i$ refer to $N,R_i^{\text{Cheb}}$ and $N^{\prime},R_i^{\text{AnMR}}$ respectively.

To reduce the second-order ODEs, we empirically find an auxiliary variable\footnote{ {Introducing $\partial_T\hat{\psi}^{[lm]}$ as an auxiliary variable will give almost the same numerical results when the \textsf{DoubleFloat64} floating-point numbers are employed.
However, the time evolution with this choice ($\partial_T\hat{\psi}^{[lm]}$) of the auxiliary variable might be unstable at late time due to increased roundoff error, if one works with double-precision floating-point numbers.}}
\begin{equation} \label{auxiliary}
    {}_{s}P^{[lm]} = C_{TT}\partial_T{}_{s}\hat{\psi}^{[lm]} + C_{TR}\partial_R{}_{s}\hat{\psi}^{[lm]} + {}_{s}C_{T}{}_{s}\hat{\psi}^{[lm]}
\end{equation}
can result in stable time evolutions at late times.
Then, the final ODEs to be solved read
\begin{equation} \label{ode}
 \begin{dcases}
	 \dfrac{\dd \vec{v}^{[l_1m_1]} }{\dd T} &={}_{-1}\bm{L}^{[l_1]} \vec{v}^{[l_1m_1]} 
	 \\
	 \dfrac{\dd \vec{v}^{[l_2m_2]} }{\dd T}  &={}_{-1}\bm{L}^{[l_2]} \vec{v}^{[l_2m_2]} 
	 \\
     \dfrac{\dd \vec{u}^{[l_3m_3]} }{\dd T}  &={}_{-2}\bm{L}^{[l_3]} \vec{u}^{[l_3m_3]} 
     \\
    & + \sum_{l_1,m_1}\sum_{l_2,m_2}
     {}_{-2}\vec{\mathcal{S}}^{[l_3m_3]}
     \left(
    \vec{v}^{[l_1m_1]};
    \vec{v}^{[l_2m_2]}
     \right) ,
\end{dcases}   
\end{equation}
where
\begin{equation}
        \vec{v} ^{[lm]} = 
        \begin{pmatrix}
        {}_{-1}\hat{\psi}^{[lm]} \\
        {}_{-1} P^{[lm]}
        \end{pmatrix}
        ,\,
        \vec{u} ^{[lm]} = 
        \begin{pmatrix}
        {}_{-2}\hat{\psi}^{[lm]} \\
        {}_{-2} P^{[lm]}
        \end{pmatrix},
\end{equation} 
and the $(2N^{(\prime)}+2)\times(2N^{(\prime)}+2)$ matrix ${}_{s}\bm{L}^{[l]}$ can be derived from the master equation \eqref{mastereq}
\begin{equation}
    {}_{s}\bm{L}^{[l]} = \begin{pmatrix}
    -{}_{s}\bm{L}_{\psi\psi}^{[l]} & \textsf{diag}[1/C_{TT}]
    \\
    -{}_{s}\bm{L}_{P\psi} & \bm{0}
    \end{pmatrix},
\end{equation}
with 
\begin{equation}
    {}_{s}\bm{L}_{\psi\psi} = 
    \textsf{diag}\left[\frac{C_{TR}}{C_{TT}}\right]\bm{D}
    +
    \textsf{diag}\left[\frac{{}_{s}C_{T}}{C_{TT}}\right]
\end{equation}
and
\begin{equation}
\begin{aligned}
{}_{s}\bm{L}_{P\psi}^{[l]} =& \textsf{diag}[C_{RR}]\bm{D}^2
    +\textsf{diag}[{}_{s}C_{R}]\bm{D}
    \\
    &+\textsf{diag}[{}_{s}C+(l-s)(l+s+1)].
\end{aligned}
\end{equation}
Here, $\textsf{diag}[C]$ denotes an $(N^{(\prime)}+1)\times(N^{(\prime)}+1)$ diagonal matrix $\bm{C}$ with diagonal elements $\bm{C}_{ii}:=C(R_i)$.

\subsection{Time-symmetric integration}
We solve the ODEs system \eqref{ode} via a fourth-order time-symmetric integration method based on the Hermite rule \cite{OBoyle2022,Markakis2023}.
The time-symmetric integration method, compared with Runge-Kutta methods of the same order, is free of Courant limit, introduces smaller truncation error, and preserves Noether charges over long time periods \cite{OBoyle2022,Markakis2023}, which makes the time-symmetric integration method ideal for long time numerical evolution in BH perturbation theory.

We will specify the time-symmetric evolution method to solve an ODEs system of the following form
\begin{equation} \label{ode_withS}
    \frac{\mathrm{d}\vec{u}}{\mathrm{d}T}=\bm{L}\vec{u}+\vec{\mathcal{S}}(T)\equiv\vec{f}(T),
\end{equation}
where $\bm{L}$ is a matrix independent of $T$ and $\vec{u}$, and $\vec{\mathcal{S}}(T)$ is a known vector function.
The problem can be converted to computing a series of integrals 
\begin{equation} \label{un2unp1}
    \vec{u}_{n+1} = \vec{u}_{n} + \int_{T_{n}}^{T_{n+1}}\vec{f}\mathrm{d}T , \quad \vec{u}_{i}:=\vec{u}(T_i),
\end{equation}
which are approximated by the Hermite rule

\begin{equation} \label{HermiteRule}
\begin{aligned}
    \int_{T_{n}}^{T_{n+1}}\vec{f}\mathrm{d}T
    =&\frac{\delta T}{2}\left(\vec{f}_{n}+\vec{f}_{n+1}\right)
    \\
    &+\frac{(\delta T)^{2}}{12}\left(\dot{\vec{f}}_{n}-\dot{\vec{f}}_{n+1}\right)+\mathcal{O}[(\delta T)^{5}]
\end{aligned}
\end{equation}
with $\delta T=T_{n+1}-T_n$.\footnote{Numerical results shown in next section \ref{sec4} are obtained with $\delta T=2^{-10}$.}
Here, the time derivative is denoted by an over-dot, and the value of a vector function $\vec{f}(T)$ at a certain time $T_i$ is denoted by a subscript $\vec{f}(T_i):=\vec{f}_i$. 
Note that eq. \eqref{un2unp1} with the Hermite rule \eqref{HermiteRule} is generally an implicit scheme to solve $\vec{u}_{n+1}$ for given $\vec{u}_{n}$. 
However, we will show that one can construct an explicit evolution scheme for ODEs of the form \eqref{ode_withS}.

The time derivatives in \eqref{HermiteRule} can be determined by
\begin{equation}
\begin{aligned}
\frac{\mathrm{d}\vec{f}}{\mathrm{d}T}&=\bm{L}\frac{\mathrm{d}\vec{u}}{\mathrm{d}T}+\frac{\mathrm{d}\vec{\mathcal{S}}}{\mathrm{d}T}.
\end{aligned}
\end{equation}
Then, \eqref{HermiteRule} becomes 
\begin{widetext}
\begin{equation}
\begin{aligned}
\int_{T_{n}}^{T_{n+1}}\vec{f}\mathrm{d}T
=&\frac{\delta T}{2}
\left[
\bm{L} \vec{u}_n + \vec{\mathcal{S}}_n
+
\bm{L} \vec{u}_{n+1} + \vec{\mathcal{S}}_{n+1}
\right]
\\
&+\frac{(\delta T)^{2}}{12}
\left[
\bm{L}^2 \vec{u}_n + \bm{L} \vec{\mathcal{S}}_n
+\dot{\vec{\mathcal{S}}}_n
-
( \bm{L}^2 \vec{u}_{n+1} + \bm{L} \vec{\mathcal{S}}_{n+1}
+\dot{\vec{\mathcal{S}}}_{n+1} )
\right]
+\mathcal{O}[(\delta T)^{5}],
\end{aligned}
\end{equation}

Thus, we obtain the time-symmetric evolution scheme for the ODEs \eqref{ode_withS} 

\begin{equation} \label{ts1}
\begin{aligned}
    \left(
    \mathbf{I} - \frac{\delta T}{2} \bm{L}
    + \frac{(\delta T)^{2}}{12}
    \bm{L}^2
    \right)
    \vec{u}_{n+1}
    = &
    \left(
    \mathbf{I} + \frac{\delta T}{2} \bm{L}
    + \frac{(\delta T)^{2}}{12}
    \bm{L}^2
    \right)
    \vec{u}_{n}
    \\
    & +
    \frac{\delta T}{2}
    \left(\mathbf{I}+\frac{\delta T}{6}\bm{L}\right)
    \vec{\mathcal{S}}_{n}
    +
    \frac{\delta T}{2}
    \left(\mathbf{I}-\frac{\delta T}{6}\bm{L}\right)
    \vec{\mathcal{S}}_{n+1}
    +
    \frac{(\delta T)^{2}}{12}
    (\dot{\vec{\mathcal{S}}}_{n} - \dot{\vec{\mathcal{S}}}_{n+1}).
    \end{aligned}
\end{equation}

Finally, we rewrite the scheme \eqref{ts1} as a form of matrix-vector multiplication and addition to reduce round-off error at each time step
\begin{equation} \label{MMA}
\begin{aligned}
\vec{u}_{n+1}=&\vec{u}_{n}+\frac{\delta T}{2}(\vec{\mathcal{S}}_{n}+\vec{\mathcal{S}}_{n+1})+\frac{(\delta T)^{2}}{12}(\dot{\vec{\mathcal{S}}}_{n}-\dot{\vec{\mathcal{S}}}_{n+1})
+
\left[\mathbf{I}-\frac{\delta T}{2}\bm{L}\left(\mathbf{I}-\frac{\delta T}{6}\bm{L}\right)\right]^{-1}(\delta T\bm{L})
\\&
\left[\vec{u}_{n}+\frac{\delta T}{3}\left(\mathbf{I}-\frac{\delta T}{8}\bm{L}\right)\vec{\mathcal{S}}_{n}+\frac{\delta T}{6}\left(\mathbf{I}-\frac{\delta T}{4}\bm{L}\right)\vec{\mathcal{S}}_{n+1}+\frac{(\delta T)^{2}}{24}\left(\mathbf{I}-\frac{\delta T}{6}\bm{L}\right)(\dot{\vec{\mathcal{S}}}_{n}-\dot{\vec{\mathcal{S}}}_{n+1})\right].
\end{aligned}
\end{equation}
\end{widetext}

For the ODEs of $\vec{v}^{[lm]}$ in \eqref{ode}, eq. \eqref{MMA} is obviously an explicit time evolution scheme.
Then, for the ODE of $\vec{u}^{[lm]}$, the source term
\begin{equation}
    \vec{\mathcal{S}}:=\sum_{l_1,m_1}\sum_{l_2,m_2}\vec{\mathcal{S}}(\vec{v}^{[l_1m_1]};\vec{v}^{[l_2m_2]})
\end{equation}
is not an explicit vector function of $T$, but $\vec{\mathcal{S}}_{n+1}(\vec{v}^{[l_1m_1]}_{n+1};\vec{v}^{[l_2m_2]}_{n+1})$
can be determined by $\vec{v}_{n}^{[lm]}$ explicitly.
In addition, the time derivative of the source term
\begin{equation}
\begin{aligned}
\frac{\mathrm{d}\vec{\mathcal{S}}}{\mathrm{d}T}
&=
\frac{\partial\vec{\mathcal{S}}}{\partial\vec{v}^{[l_1m_1]}}
\frac{\mathrm{d}\vec{v}^{[l_1m_1]}}{\mathrm{d}T}
+\frac{\partial\vec{\mathcal{S}}}{\partial\vec{v}^{[l_2m_2]}}
\frac{\mathrm{d}\vec{v}^{[l_2m_2]}}{\mathrm{d}T} 
\\
&=
\frac{\partial\vec{\mathcal{S}}}{\partial\vec{v}^{[l_1m_1]}}
{}_{-1}\bm{L}^{[l_1]}\vec{v}^{[l_1m_1]}
+
\frac{\partial\vec{\mathcal{S}}}{\partial\vec{v}^{[l_2m_2]}}
{}_{-1}\bm{L}^{[l_2]}\vec{v}^{[l_2m_2]}
\end{aligned}
\end{equation}
can also be determined by $\vec{v}_{n}^{[lm]}$ explicitly.
Hence, we obtain an explicit time evolution scheme for the whole ODEs system \eqref{ode}.

\section{Numerical results} \label{sec4}

In this section, we discuss choices of initial data and mode coupling first.
Then we present our main result, i.e., the power law of second-order tails, followed by its implications for multi-messenger observations.
Finally, we describe numerical checks of our numerics. 

\subsection{Initial data and mode coupling}
It is known that the decay rates of linear tails depend on initial data, e.g., the angular profile, the time derivative of the field (stationary or not) or the radial profile (compact support or not) \cite{Poisson2002,Price2004,Burko2011,Harms2013}.

For the first-order electromagnetic perturbation $\phi_2^{(1)}$, we employ the initial data that are commonly used in the literature.

Specifically, the angular profile is described by spin-weighted spherical harmonics with a single multipole $l$ for simplicity, i.e.,
\begin{equation} \label{pure_multipole}
\begin{aligned}
\phi_2^{(1)}(T=0,R,\theta,\varphi) 
&=\sum_{m}{}_{-1}\hat{\psi}^{[lm]}{}_{-1}Y_{lm}
\\
&={}_{-1}\hat{\psi}^{[l]}{}_{-1}P_{l}\sum_{m}\ee^{\ii m \varphi}
\\
\text{with } {}_{-1}Y_{lm}(\theta,\varphi) &= {}_{-1}P_{l}(\theta)\ee^{\ii m \varphi},
\end{aligned}
\end{equation}
where we assume ${}_{-1}\hat{\psi}^{[lm]}\equiv{}_{-1}\hat{\psi}^{[l]}$ for different azimuthal numbers $m$.

For the time derivative of the field, we only consider the so-called stationary initial data, i.e.,
\begin{equation} 
     {}_{-1}\hat{\psi}^{[l]}_{,T}(T=0,R) \equiv 0.
    \label{dpsiinit}
\end{equation}
Note that there is no regular solution $\psi(R)$ to an ODE $C_{TR}\pd_R\psi+{}_{s}C_{T}\psi=0$, for which the auxiliary variable ${}_{-1}P^{[l]}$ \eqref{auxiliary} does not vanish initially.
Consequently, the power law of ${}_{-1}\hat{\psi}^{[l]}$ is determined by the late-time behavior of the retarded Green's function $\mathcal{G}(T-T^\prime,R,R^\prime)$, following from the homogeneous solution\footnote{We also verified that the power law, with the initial data such that ${}_{s}P^{[l]}(T=0,R)\equiv0$, is one less than that obtained with the so-called stationary initial data \eqref{dpsiinit}.
This indicates that it is governed by the late-time behavior of $ \mathcal{G}_{,T^\prime}$.
In this sense, it is more appropriate to refer to  ${}_{s}P^{[l]}(T=0,R)\equiv0$ as ``stationary initial data''.}
\begin{widetext}
\begin{equation}\label{homo}
\begin{aligned}
    {}_{s}\hat{\psi}^{[l]}(T,R)
    =&\int_{0}^{R_\text{H}}
    \left\{
    {}_{s}P^{[l]}(T^\prime,R^\prime)
    \mathcal{G}(T-T^\prime,R,R^\prime)
    -
    C_{TT}(R^\prime)
    {}_{s}\hat{\psi}^{[l]}(T^\prime,R^\prime)
    \mathcal{G}_{,T^\prime}(T-T^\prime,R,R^\prime)
    \right\}_{T^\prime=0}
    \mathrm{d}R^\prime
\end{aligned}
\end{equation}
\end{widetext}
to the (1+1)-dimensional PDE \eqref{mastereq} with ${}_s\slashed{\Delta}$ being replaced by the eigenvalue $-(l-s)(l+s+1)$, where $\mathcal{G}(T-T^\prime,R,R^\prime)$ is a solution to the PDE with a Dirac-delta source term $\delta(T-T^\prime)\delta(R-R^\prime)$ and is subject to the condition that $\mathcal{G}(T-T^\prime,R,R^\prime)=0$ for $T<T^\prime$.

The radial profile is given by
\begin{align}
    {}_{-1}\hat{\psi}^{[l]}(T=0,R) &= G(R;k,b) ,
    \label{psiinit}
\end{align}
with
\begin{equation}
    G(R;k,b) = k\exp\left[-\left(\frac{R-R_c}{w}
    \right)^2\right]+b.
\end{equation}
The results shown in this work are obtained with $R_c=R_\text{H}/2$ and $w=R_\text{H}/10$.

In this paper, we refer to the initial data \eqref{psiinit} with
\begin{itemize}
    \item $(k,b) = (1,0)$ as compact support, \footnote{Even if a Gaussian wave packet is not strictly compact support and the boundary value $G(R=0;1,0)=\ee^{-(R_0/w)^2}=\ee^{-25}\simeq1.4\times10^{-11}$ is much larger than the round-oﬀ error of the \textsf{DoubleFloat64} floating-point numbers $2^{-104}\simeq4.9\times10^{-32}$, {the decay rates of linear tails for compact support initial data are successfully reproduced (see Fig. \ref{LPIphi}).}}
    \item $(k,b) = (0,1)$ as non-compact support.
\end{itemize}  

These types of initial data are usually employed to study linear tails.
Besides, we also take initial data employed in the literature concerned with nonlinear tails \cite{Cardoso2024,Ling2025}, i.e., ingoing data $(\Delta{}_{-1}\hat{\psi}^{[l]}=0$),
\begin{align}
    {}_{-1}\hat{\psi}^{[l]}(T=0,R) &= G(R;k=1,b=0) ,
    \\
     {}_{-1}\hat{\psi}^{[l]}_{,T}(T=0,R) &= -\frac{n^R}{n^T}
    \partial_R
    {}_{-1}\hat{\psi}^{[l]}(T=0,R), 
\end{align}
and outgoing data $(D{}_{-1}\hat{\psi}^{[l]}=0)$,
\begin{align}
    {}_{-1}\hat{\psi}^{[l]}(T=0,R) &= G(R;k=1,b=0) ,
    \\
  {}_{-1}\hat{\psi}^{[l]}_{,T}(T=0,R) &= -\frac{l^R}{l^T}
    \partial_R
    {}_{-1}\hat{\psi}^{[l]}(T=0,R).
\end{align}

The above types of initial data, plus zero initial data 
\begin{equation}
    {}_{-2}\hat{\psi}^{[l]}(T=0,R)=
    {}_{-2}\hat{\psi}^{[l]}_{,T}(T=0,R) \equiv0,
\end{equation}
are employed for the second-order gravitational perturbation $\Psi_4^{(2)}$.

The pure multipole initial data \eqref{pure_multipole} simplify the mode coupling, as the coupling of $l$ and $m$ in the source terms \cref{S1,S2,S3} can be separated.
Different mode coupling channels $(m_1,m_2)$ such that $m_1-m_2=m_3$ just differ by a constant \footnote{Thus, we do not find {exceptions, e.g., the coupling of $l=m=2$ to $l=-m=2$ reported in \cite{Cardoso2024}}, to the power law \eqref{ALPI} described in Sec. \ref{sec_seclaw}.}
 {
\begin{equation}
    (-1)^{m_2}
    \begin{pmatrix}
    l_1&l_2&l_3
    \\
    m_1&-m_2&-m_3
    \end{pmatrix}
\end{equation}
in the source terms.}
Thus, we only need to consider mode coupling of $(l_1=l,l_2=l)$ such that $l_3=2,3,\cdots,2l$ for the pure multipole case.

We checked that the power law of second-order tails does not depend on the mode coupling channels $(l_1,l_2)\to l_3$ and that it also does not depend on the initial data of ${}_{-1}\hat{\psi}^{[l]}$ and ${}_{-2}\hat{\psi}^{[l_3]}$ both when $l_3\geq4$.

\subsection{The power law of second-order tails} \label{sec_seclaw}
The power law decay of a field $\psi$ is monitored by local power index (LPI)
 {
\begin{equation}
    \text{LPI} = \frac{\partial\log|\psi|}{\partial\log T},
\end{equation}
}
such that the LPI is a constant $p$ if the field $\psi\sim T^{p}$.

\begin{figure*}[htp]
    \centering
    \includegraphics[width=5.4cm]{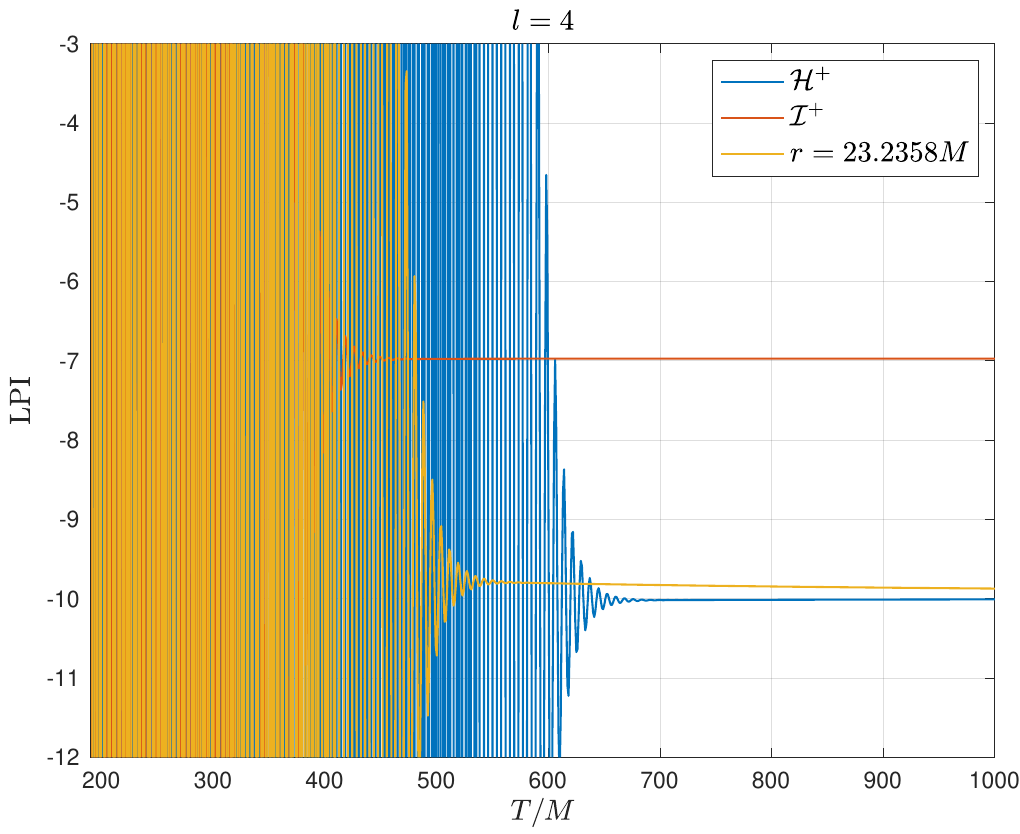}
    \includegraphics[width=5.4cm]{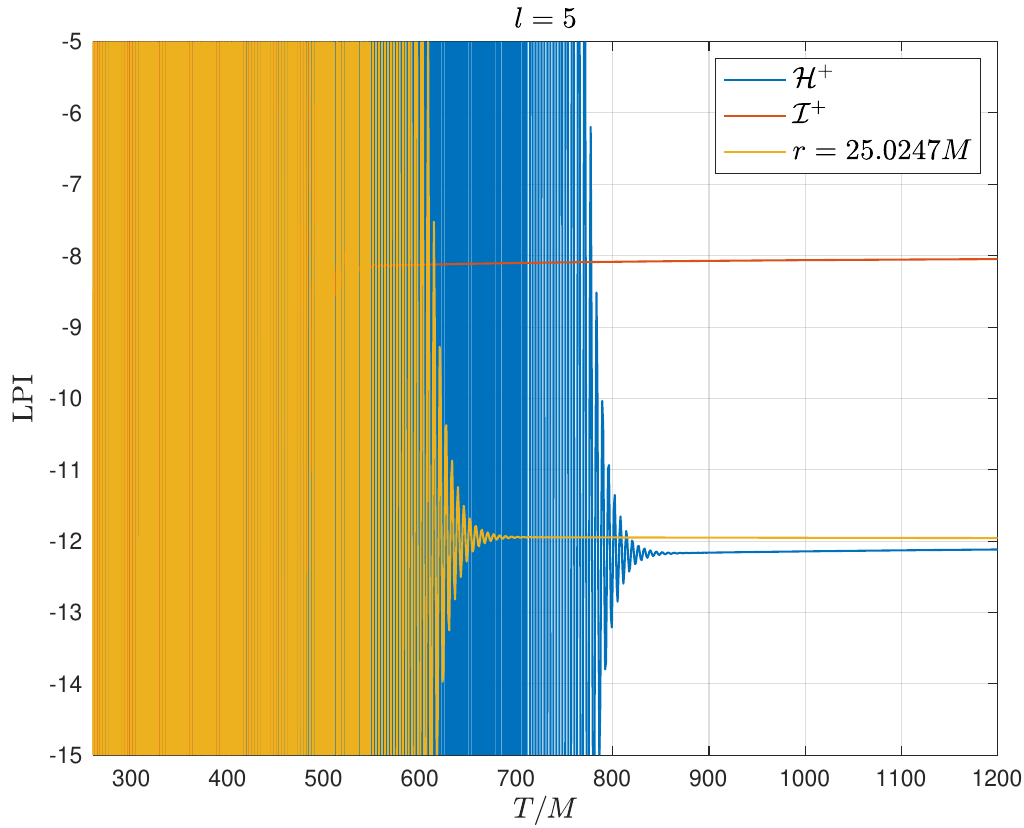}
    \includegraphics[width=5.4cm]{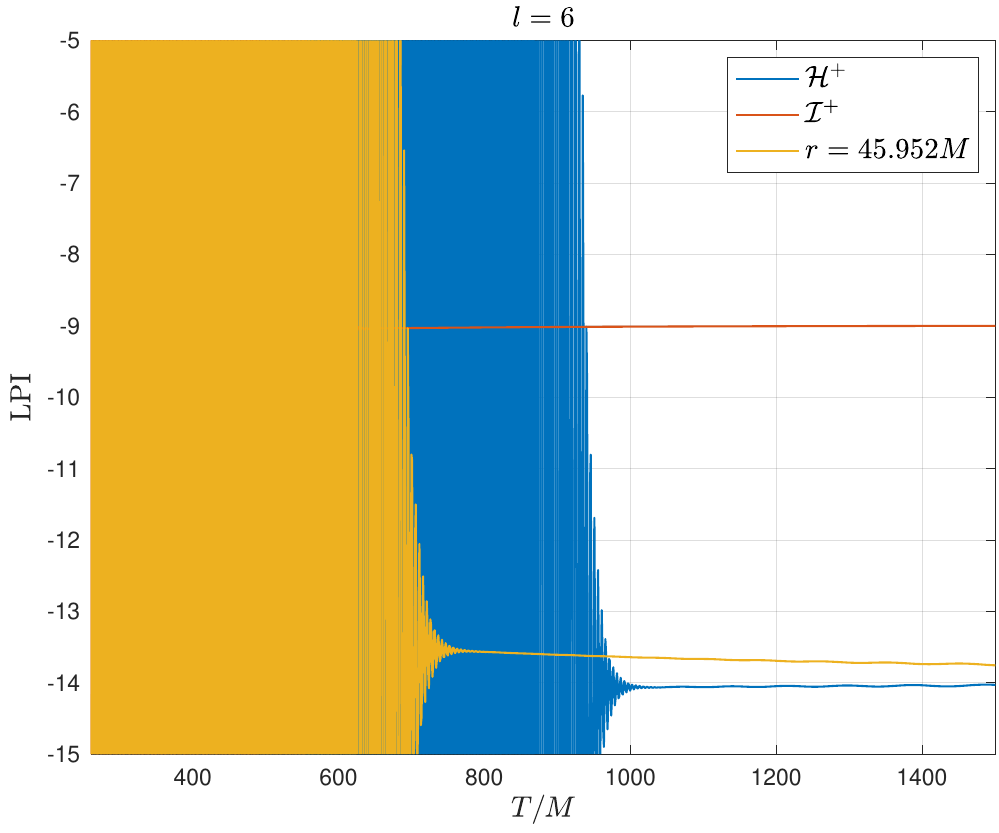}
    \caption{LPIs of second-order perturbation ${}_{-2}\hat{\psi}^{[l]}$ for $l=4,5,6$ suggest a power law of the form $T^{-2l-2}$ at $\mathcal{H}^+$ and finite radii, and $T^{-l-3}$ at $\mathcal{I}^+$.
    }
    \label{LPIpsi4}
\end{figure*}

\begin{figure}[htp]
    \centering
    \includegraphics[width=0.48\textwidth]{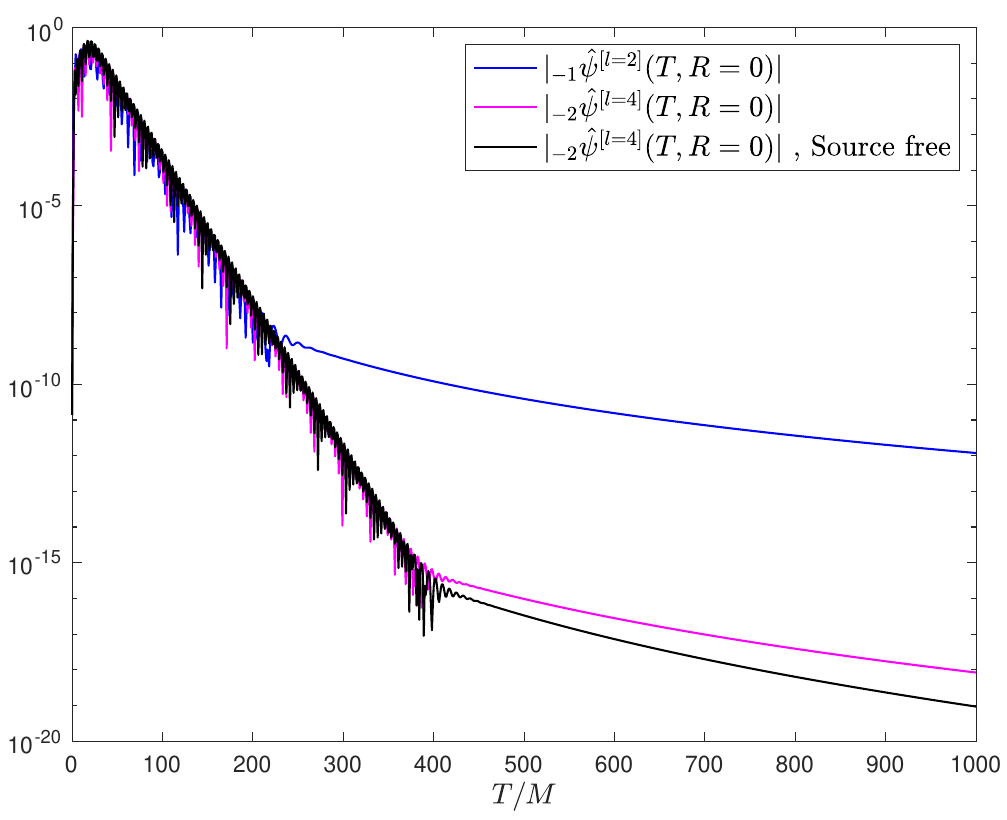}
    \caption{Waveform of gravitational perturbation ${}_{-2}\hat{\psi}$ and its parent electromagnetic perturbation ${}_{-1}\hat{\psi}$, as well as a gravitational perturbation ${}_{-2}\hat{\psi}$ solved by the source-free BPT equation.
    Here, the same ingoing initial data are set for all the perturbative variables.
    The source-driven tail dominates over the source-free tail, supporting the breakdown of linear perturbation theory at late times.}
    \label{logWF}
\end{figure}

Our main results are shown in the Fig. \ref{LPIpsi4}.
The LPIs of ${}_{-2}\hat{\psi}^{[l]}$ for $l=4,5,6$ suggest a power law of the form 
\begin{equation} \label{NLPI}
    \begin{dcases}
    T^{-2l-2}&,\,\text{at fixed spatial position,}
    \\
    T^{-l-3}&,\,\text{at future null infinity},
\end{dcases} 
\end{equation}
and this power law, as mentioned above, is independent of the initial data, indicating that in the case $l\geq4$ the homogeneous part \eqref{homo} of ${}_{-2}\hat{\psi}^{[l]}$ cannot dominate over its inhomogeneous part
\begin{equation}\label{inho}
    \int_{0}^T\int_{0}^{R_\text{H}} \mathcal{G}(T-T^\prime,R,R^\prime)
    {}_{-2}\hat{\mathcal{S}}(T^\prime,R^\prime)
    \mathrm{d}R^\prime
    \mathrm{d}T^\prime.
\end{equation}
This result supports a dominant position of nonlinear tails over their linear counterparts, in line with \cite{Cardoso2024,Ling2025,Kehagias2025}.
To see this more clearly, we solve both the inhomogeneous and homogeneous (namely ${}_{-2}\hat{\mathcal{S}}=0$) BPT equations \eqref{mastereq} under the same initial data.
As shown in Fig. \ref{logWF}, the source-driven perturbation overshadows the source-free one at late times, thereby determining how a perturbed BH approaches a stationary state and playing a critical role in fundamental theoretical issues, e.g., strong cosmic censorship. \cite{Cardoso2018,Cardoso2018a,Mo2018,Liu2019,Hod2019,Zhang2019,Luna2019,Dias2019}.

\begin{figure*}[htp]
    \centering
    \includegraphics[width=8.1cm]{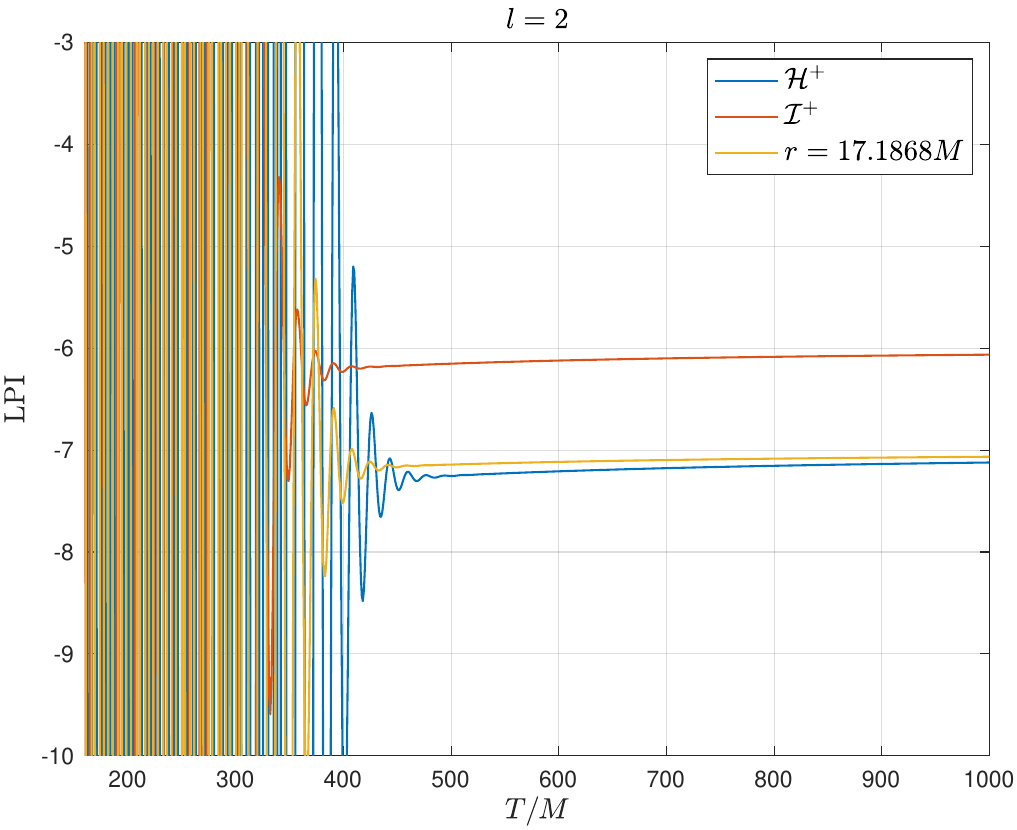}
    \includegraphics[width=8.1cm]{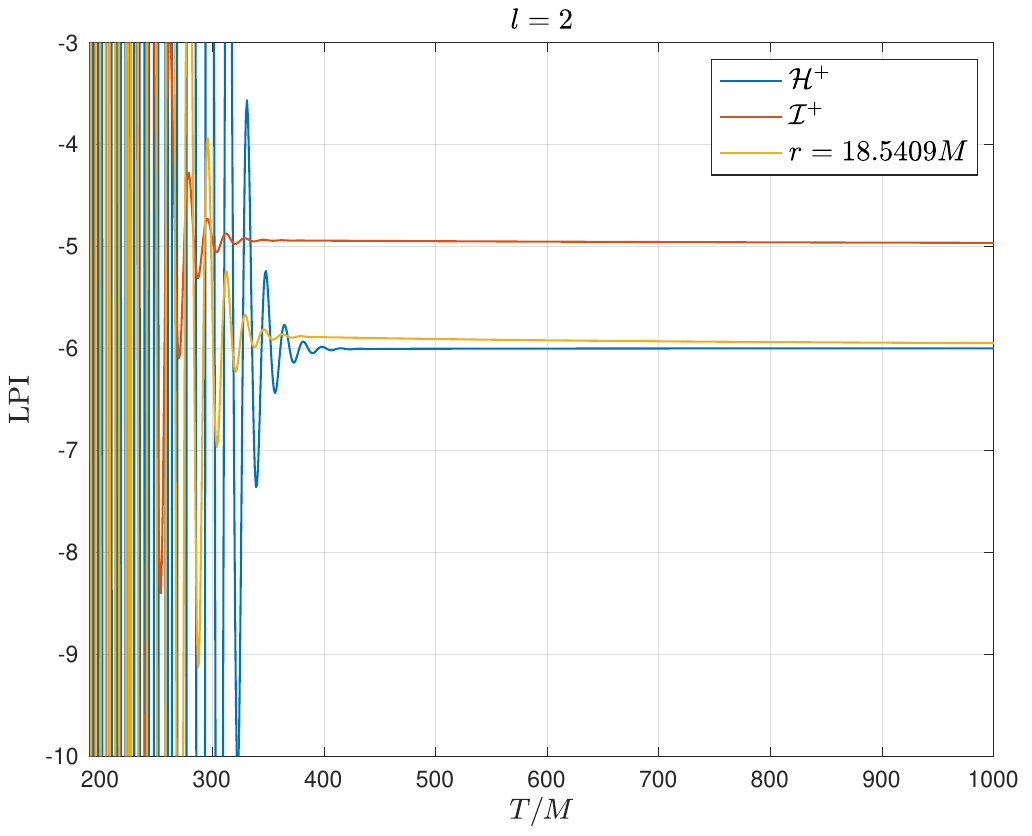}
    \includegraphics[width=8.1cm]{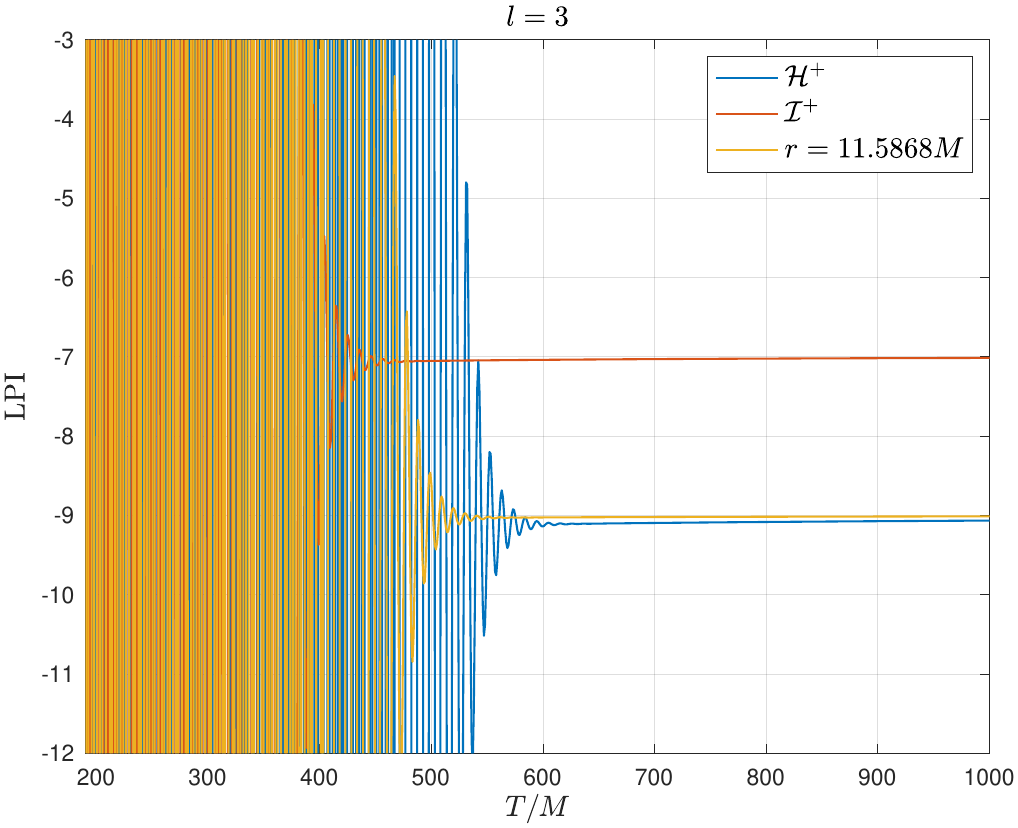}
    \includegraphics[width=8.1cm]{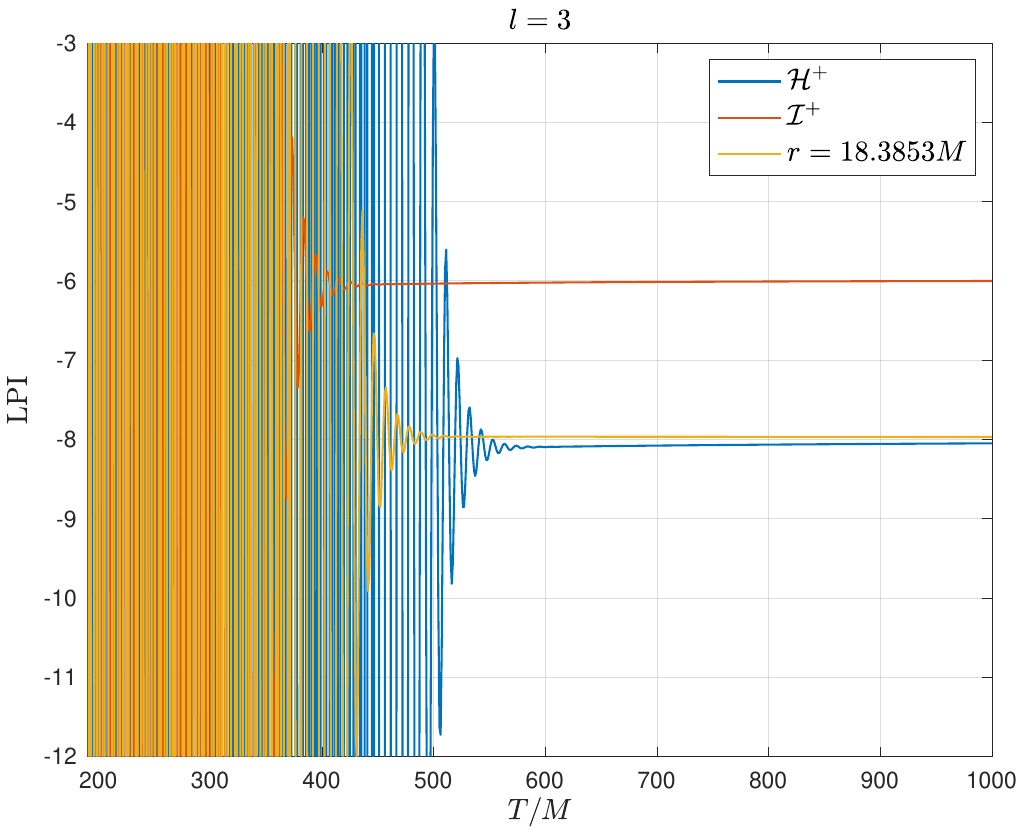}
    \caption{LPIs of second-order perturbation ${}_{-2}\hat{\psi}^{[l]}$ when $l=2$ (upper panels) and $l=3$ (lower panels) for ${}_{-1}\hat{\psi}$ with the compact (left panels) and non-compact (right panels) support initial data, extracted at future null infinity $\mathcal{I}^+$, future event horizon $\mathcal{H}^+$ and finite radii, respectively.
    }
    \label{LPIpsi4l23}
\end{figure*}

When $l=2$ and $3$, however, the power law of second-order perturbation ${}_{-2}\hat{\psi}^{[l]}$ depends on the initial data of ${}_{-1}\hat{\psi}$ and ${}_{-2}\hat{\psi}^{[l]}$ both.
Interested in the effect of source term, we focus on results with the zero initial data for ${}_{-2}\hat{\psi}^{[l]}$, as shown in Fig. \ref{LPIpsi4l23}.
We find that for ${}_{-1}\hat{\psi}$ with the compact support initial data, ${}_{-2}\hat{\psi}^{[l]}$ decays as $T^{-2l-3}$ at fixed spatial position and $T^{-l-4}$ at future null infinity $\mathcal{I}^+$, which are the same as the linear Price's law.
However, the non-compact support initial data of ${}_{-1}\hat{\psi}$, forming an extended source at the beginning, lead to a slower tail that conforms to the power law \eqref{NLPI}.

Recent analytical results \cite{Cardoso2024,Ling2025,Kehagias2025}, formulated in the Regge-Wheeler-Zerilli (RWZ) formalism, indicate that the behavior of the source term $Q$ at large radii $Q\sim 1/r^\beta$ plays a key role in determining the power law of nonlinear tails. 
It was shown that the inhomogeneous part of Regge-Wheeler/Zerilli master function $\Psi$, induced by a compact outgoing source term $Q\sim 1/r^\beta$, decays as  
\begin{equation} \label{ALPI}
    \begin{dcases}
    t^{-\beta-l} &\text{for } \beta =0,1,
    \\
    t^{-2l-2} &\text{for } 2\leq\beta\leq l+2,
    \\
    t^{-2l-3} &\text{for } \beta\geq l+3,
    \end{dcases}
\end{equation}
at fixed spatial position \cite{Ling2025}. \footnote{The homogeneous part of $\Psi$, concerned with the initial data, was ignored in \cite{Ling2025}.}
Note that the curvature perturbation $\Psi_4$ is related to the metric perturbation $\Psi$ by the Chandrasekhar transformation \cite{Chandrasekhar1998,Lousto2005}. 
The decay rates of $\Psi_4$ and $\Psi$ are the same near the black hole, but the former is two less than the later at null infinity \cite{Zenginoglu2010}.
Our numerical result, i.e., the power law $T^{-2l-2}$ at fixed spatial position when $l\geq4$ for compact support ${}_{-1}\hat{\psi}(T=0,R)$, coincides with the analytical prediction \eqref{ALPI} for the case $\beta=6$. \footnote{See \cite{Aly2025} for a derivation of the electromagnetic source in RWZ formalism.}
In addition, to our knowledge, the dependence of the second-order power law on the initial data of the first-order quantities had not been discovered before.
However, the power law $T^{-l-3}$ at future null infinity $\mathcal{I}^+$, or $T^{-l-1}$ in the metric perturbation picture, is inconsistent with the analytical prediction 
\begin{equation}
    \begin{dcases}
    u^{-\beta} &\text{when } 2\leq\beta\leq l+1 \text{ for } l\geq1
    \\
    u^{1-\beta} &\text{otherwise}
    \end{dcases}
\end{equation}
presented in \cite{Cardoso2024}, \footnote{But our finding is consistent with their numerical result of second-order tails for the self-coupling of $l=m=2$.} which means that their analytical approach omits some features \cite{Ling2025}.

\subsection{Implications for multi-messenger observations}
Despite still being strongly suppressed by the QNMs and neglected by any near-future observatories, the nonlinearly source-driven tails can be significantly amplified for binary BHs with high eccentricities, prospectively reaching the detection threshold of upcoming detectors \cite{Albanesi2023,Carullo2023,DeAmicis2024a,Islam2025}.
Hence, it would be valuable to discuss the implications of our results for multi-messenger observations in this subsection.

Foremost, the electromagnetically sourced nonlinear tails differ from gravitationally sourced nonlinear tails when $l=2,3$, because the latter, with a source $Q\sim1/r^2$ \cite{Brizuela2006,Nakano2007}, decay as $T^{-2l-2}$ at fixed spatial position when $l\geq2$ even if the first-order quantities are initially compact support \cite{Cardoso2024}\footnote{We conjecture that the gravitationally sourced nonlinear tails decay as $u^{-l-1}$ in the metric perturbation picture, albeit this was not explicitly reported in \cite{Cardoso2024}.}.
This difference, combined with the analysis of quadratic QNM \cite{Jana2024}, could help to identify astrophysical origins of GWs in the multi-messenger observations and offers a novel and complementary mechanism to detect black holes in the Milky Way.

\begin{figure*}[htp]
    \centering
    \includegraphics[width=8.1cm]{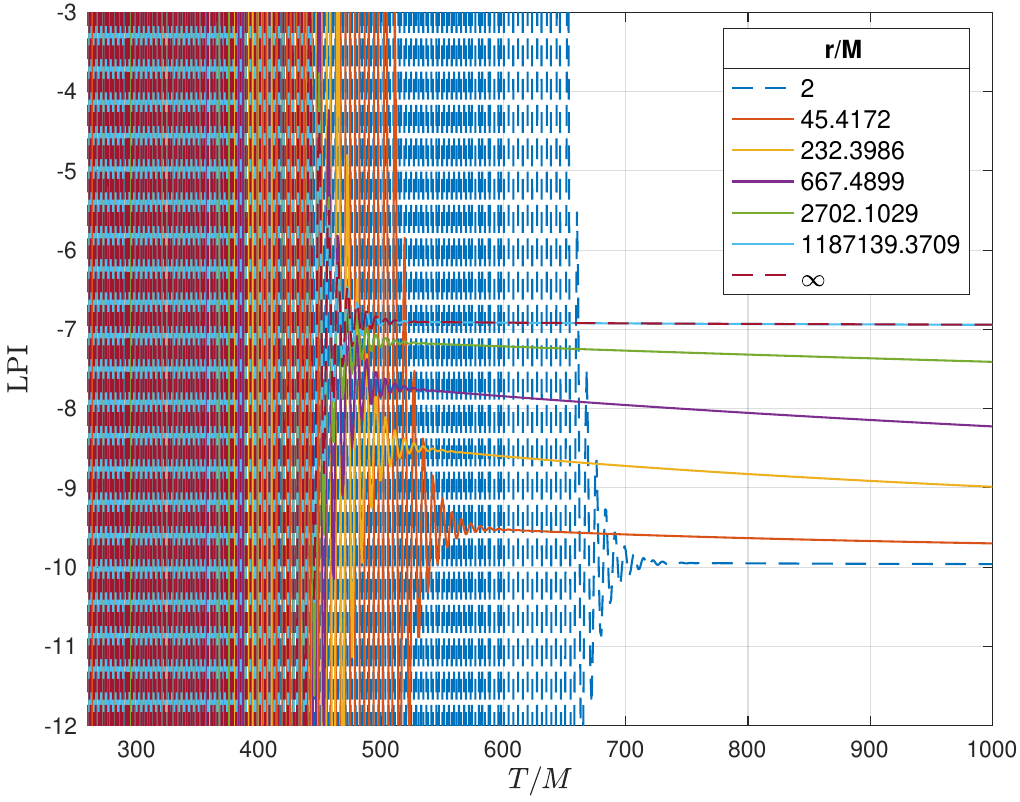}
    \includegraphics[width=8.1cm]{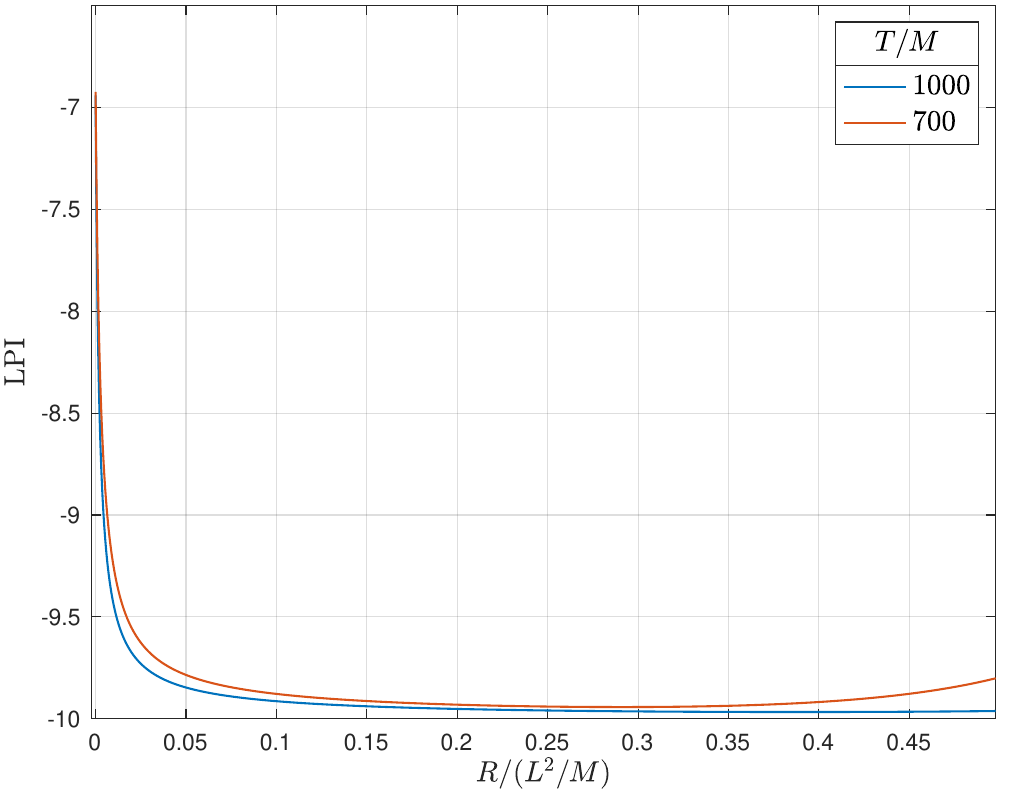}
    \caption{Distance-dependence of decay rates $p(T,R)$ for the second-order tails.
    The point closest to the future null infinity $\mathcal{I}^+$ in the numerical grid is located at $L^2/R^{\text{AnMR}}_{N^\prime-1}=1187139.3709M\simeq1.2\times10^6M$, where the decay rate is almost the same as that at $\mathcal{I}^+$ during the whole evolution.
    Therefore, the curves of the two are almost identical in the left panel.
    For comparison, the closest candidate for a supermassive black hole, Sgr A$^*$, with a mass $M=3.7\times10^{6}M_{\odot}$ is about 26000 light years away, which roughly corresponds to $1.8\times10^9M$ in the geometric units.}
    \label{pointLPI}
\end{figure*}

As depicted in Fig. \ref{pointLPI}, the LPIs of second-order tails show a similar distance-dependence with linear tails \cite{SZPAK2008,SZPAK2009,Purrer2005,Zenginoglu2008a}, i.e., the decay rates $p(T,R)$ at finite distance vary monotonously between that at null infinity $\mathcal{I}^+$ and that at event horizon $\mathcal{H}^+$.
However, it is the decay rate at null infinity $\mathcal{I}^+$ that is relevant for astronomical observations, due to extremely distant astronomical distances \cite{Zenginoglu2008a}.

\begin{figure*}[htp]
    \centering
    \includegraphics[width=8.1cm]{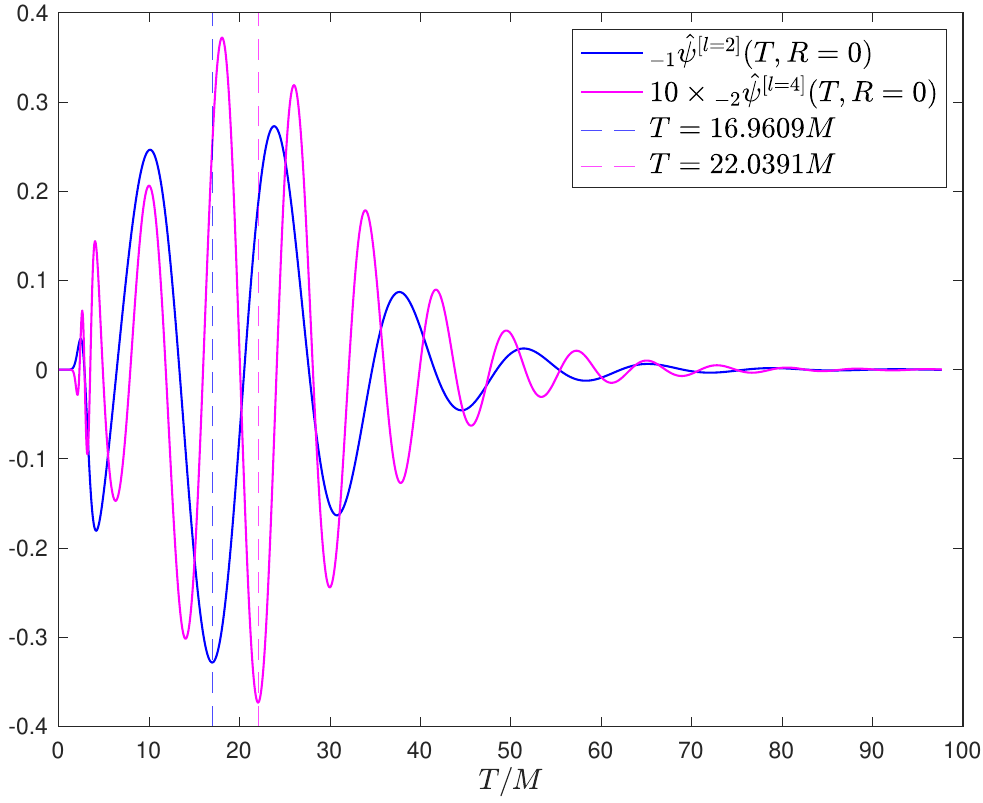}
    \includegraphics[width=8.1cm]{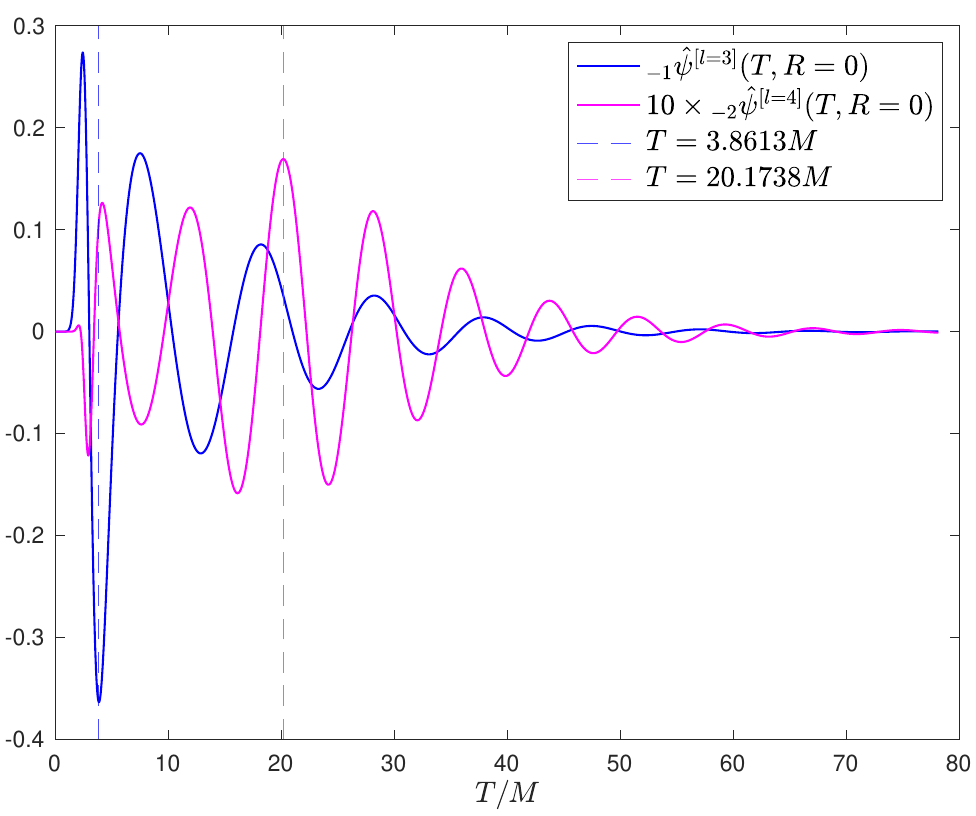}
    \caption{Waveform of gravitational perturbation ${}_{-2}\hat{\psi}$ and its parent electromagnetic perturbation ${}_{-1}\hat{\psi}$ extracted at null infinity $\mathcal{I}^+$.
     {In the left (right) panel, we set the ingoing (outgoing) initial data for ${}_{-1}\hat{\psi}^{[l=2\,(3)]}$ and set the zero initial data for ${}_{-2}\hat{\psi}^{[l=4]}$.}
    Note that the values of ${}_{-2}\hat{\psi}$ have been enlarged in the figure.
    The peaks of the waveform are denoted by vertical dashed lines.
    In our simulations, the peaks of ${}_{-1}\hat{\psi}$ typically arrive at $\mathcal{I}^+$ first.
    In the figure, the time difference $\Delta T\sim10M$ of ${}_{-1}\hat{\psi}$'s peak and ${}_{-2}\hat{\psi}$'s peak roughly corresponds to $10^3$ sec., if we set $M=10^6M_{\odot}$.
    }
    \label{waveform}
\end{figure*}

As shown in Fig. \ref{waveform}, both the evolutions of ${}_{-1}\hat{\psi}$ and ${}_{-2}\hat{\psi}$ consist of three distinct phases: initial transient, quasi-normal ringing, and polynomial tail decay.
During the quasi-normal ringing phase, we extract the quasi-normal frequencies with the matrix pencil method \cite{Berti2007}.
The spectrum of the first-order quantities ${}_{-1}\hat{\psi}$ matches the prediction of linear BH perturbation theory.
Besides the linear QNMs, the quadratic QNMs with frequencies twice that of ${}_{-1}\hat{\psi}$'s linear mode are also found, which come from the free propagation part $\mathcal{G}_F$ of the Green’s function $\mathcal{G}$ in \eqref{inho} \cite{Okuzumi2008}.
Moreover, the peaks of GW waveform typically arrive at $\mathcal{I}^+$ later than those of their parent EM waveform in our simulations, indicating that EM events could be a forecast of their offspring GW events.
These timing and spectral signatures provide practical search priors for multi-messenger campaigns: the EM peak time can trigger time-gated and stacked searches for long-lived ringdown tails in GW data, thereby lowering detection thresholds \cite{Kelley2013,Yang2017,Abbott2023}.

\subsection{Numerical checks}\label{sub_NCheck}
\begin{figure*}[htp]
    \centering
    \includegraphics[width=5.4cm]{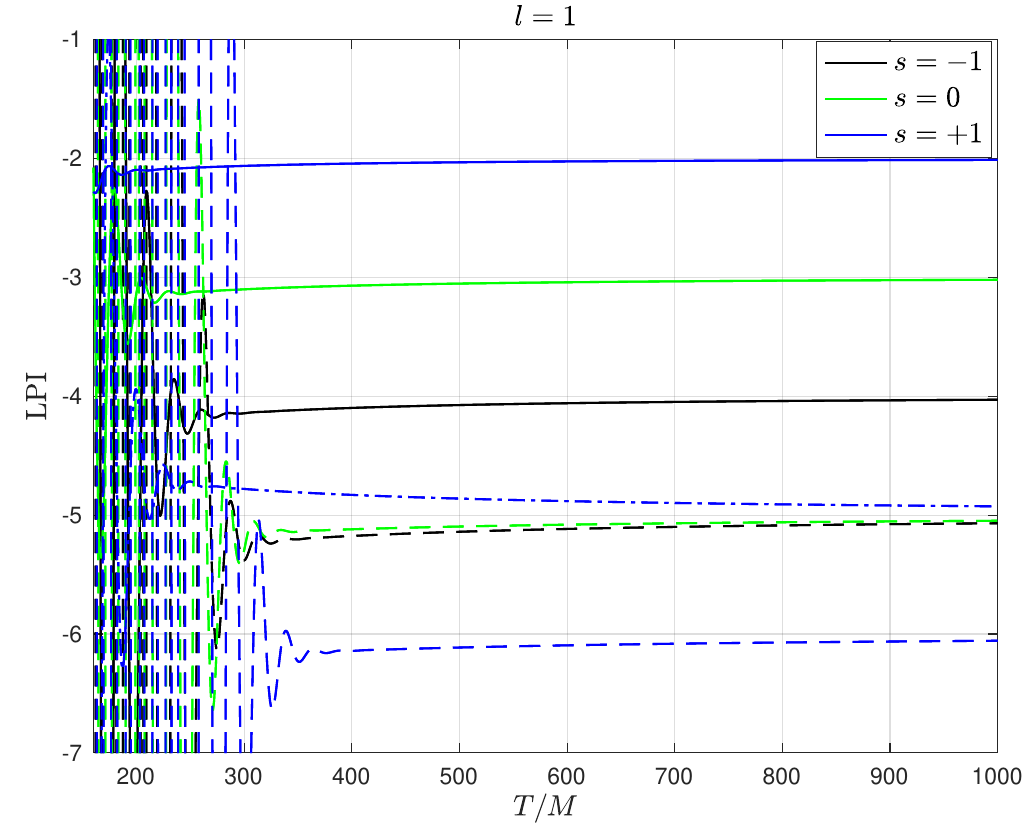}
    \includegraphics[width=5.4cm]{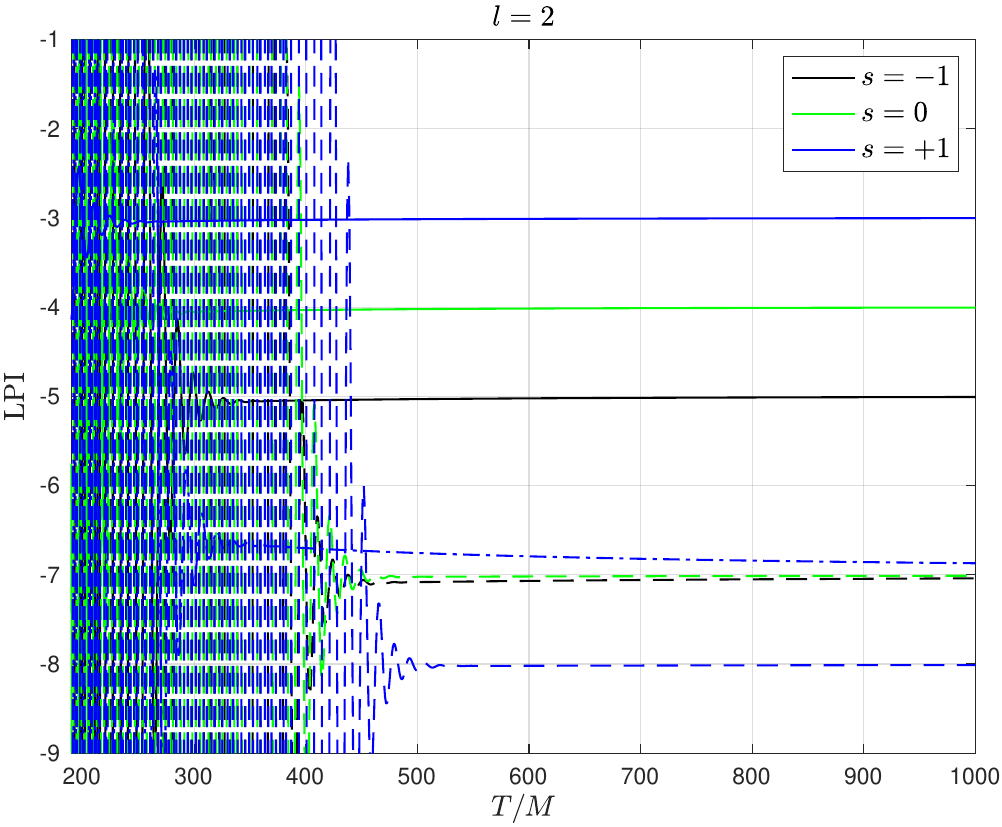}
    \includegraphics[width=5.4cm]{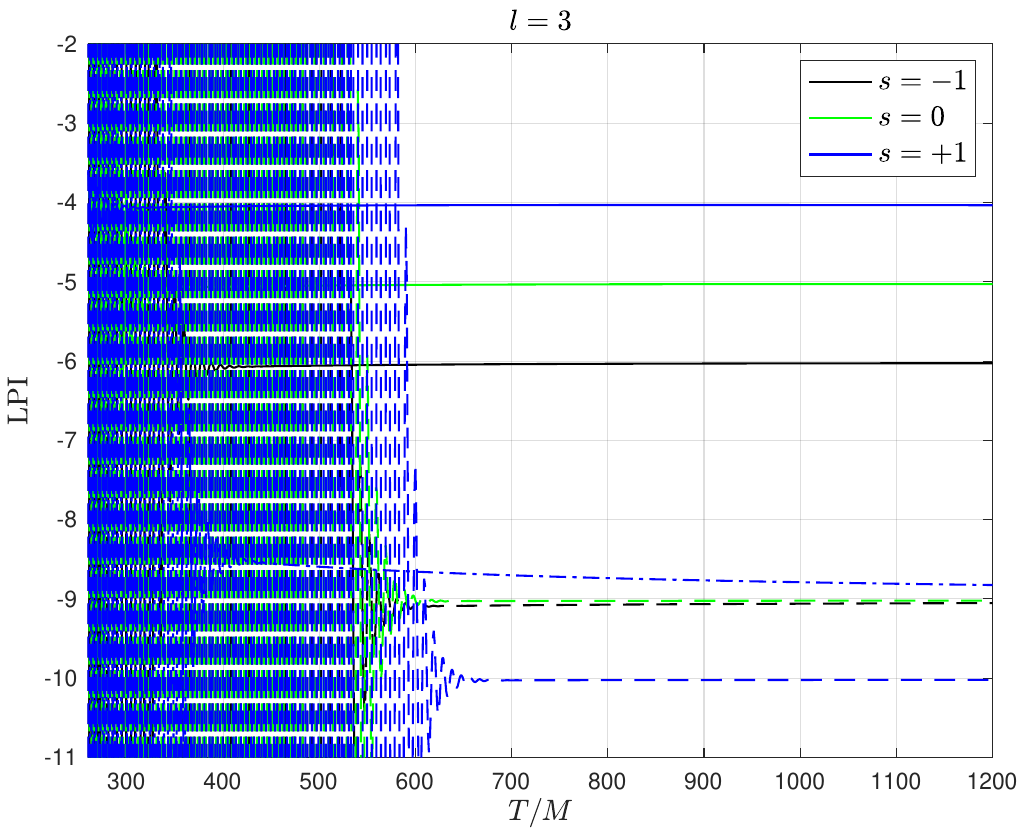}
    \caption{The LPIs of $\phi_2^{(1)}\,(s=-1)$ , $\phi_1^{(1)}\,(s=0)$ and $\phi_0^{(1)}\,(s=+1)$ for $l=1,2,3$ from left to right.
    Here, the compact support initial data are employed.
    Solid, dashed, and dashed-dotted lines represent LPIs that are extracted at future null infinity $\mathcal{I}^+$, future event horizon $\mathcal{H}^+$, and finite radii, respectively.
    The power law of $\phi_1^{(1)}$ shown here is consistent with that obtained by solving the Fackerell-Ipser equation \cite{Racz2024}.
    The LPI splitting for $\phi_0^{(1)}$ is successfully reproduced.
    We also checked that the power law of $\phi_0^{(1)}$ for non-compact support initial data, not shown here, accords with \cite{Macedo2014}.}
    \label{LPIphi}
\end{figure*}

Our numerics are checked by successful reproduction of the power law for the first-order Maxwell scalars $\phi_2^{(1)},\phi_1^{(1)}$ and $\phi_0^{(1)}$ (see Fig. \ref{LPIphi}), as well as spectral accuracy of evolutionary variables (see Fig. \ref{spec_end}).

An accurate reconstruction of $\phi_0^{(1)}$ requires quite a few collocation points, due to a LPI splitting\footnote{The LPI splitting is that a field with positive spin weight decays with three different rates at $\mathcal{H}^+,\mathcal{I}^+$ and the bulk \cite{Hod2000a,Barack1999a}.} and a greater difference of LPIs between $\mathcal{I}^+$ and finite radii.
However, we find that even if the number of collocation points is not large enough to reproduce the power law of $\phi_0^{(1)}$ at late times, the power law of ${}_{-2}\hat{\psi}$ is not affected.

\begin{figure}[htp]
    \centering
    \includegraphics[width=0.48\textwidth]{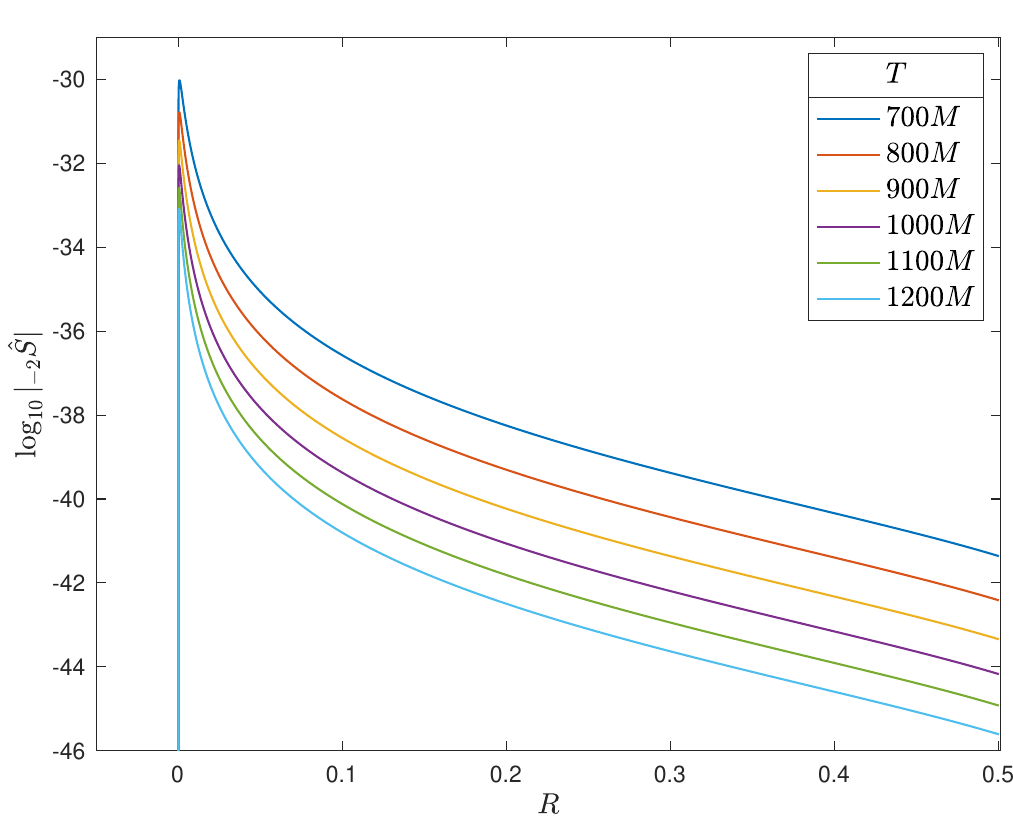}
    \caption{Profiles of rescaled source term ${}_{-2}\hat{\mathcal{S}}^{[55]}({}_{-1}\hat{\psi}^{[33]};{}_{-1}\hat{\psi}^{[3,-2]})$ at different time slices.
    Note that the ordinate is logarithmic.
    The source term resembles a Dirac-delta distribution in the compactified $R$ coordinate.}
    \label{logS}
\end{figure}

\begin{figure}[htp]
    \centering
    \includegraphics[width=0.5\textwidth]{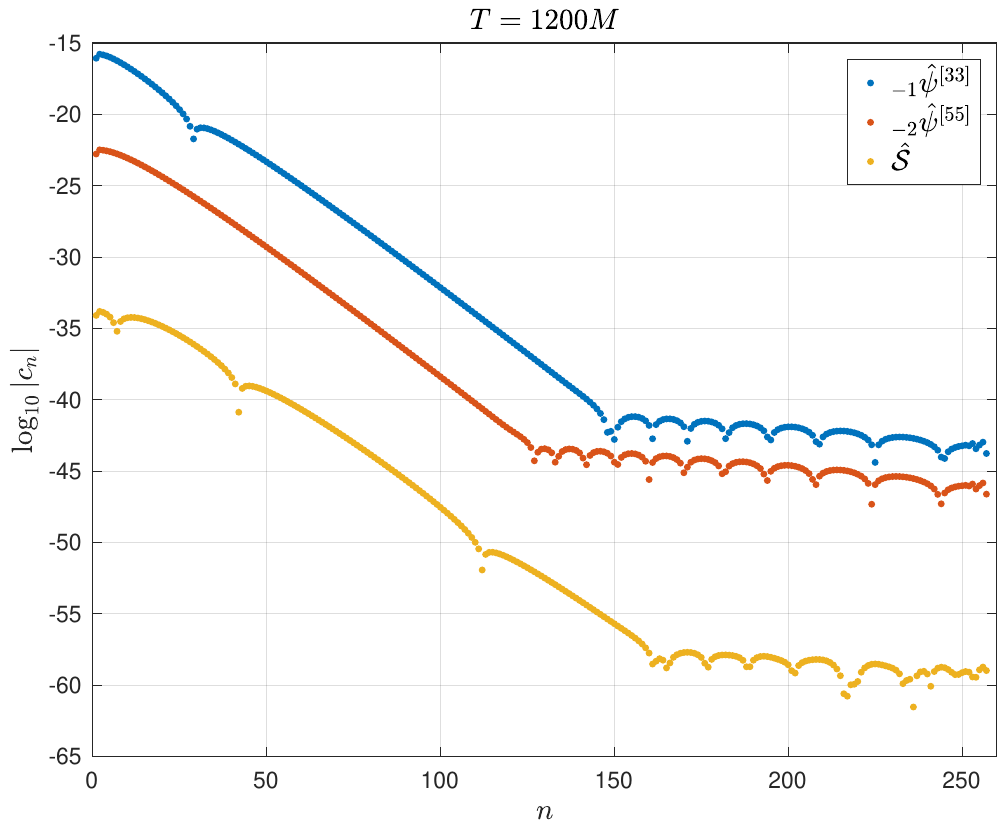}
    \caption{Spectral convergence of  ${}_{-2}\hat{\mathcal{S}}$, ${}_{-1}\hat{\psi}$ and ${}_{-2}\hat{\psi}$ at $T=1200M$.}
    \label{spec_end}
\end{figure}

The rescaled source term ${}_{-2}\hat{\mathcal{S}}^{[lm]}$ \eqref{Sres}, shown in Fig. \ref{logS}, resembles a Dirac-delta distribution in the compactified $R$ coordinate, whose peak approaches $R=0$ at a coordinate speed much less than the outgoing characteristic $c_+(R=0)=-1/8$ (see Table \ref{S_peak}). 
This Dirac-delta-like profile can severely spoil the convergence of spectral methods.
Nevertheless, with the AnMR collocation points \eqref{AnMR}, spectral coefficients of the source term ${}_{-2}\hat{\mathcal{S}}$, as well as those of the evolutionary variables ${}_{-1}\hat{\psi}$ and ${}_{-2}\hat{\psi}$, indicate an exponential convergence rate of the spectral expansions (see Fig. \ref{spec_end}).
In addition, no obvious aliasing is found in the nonlinear source term ${}_{-2}\hat{\mathcal{S}}$, for which we do not apply any filter in our evolutions.

\begin{table}
    \centering
    \begin{tabular}{c|c|c|c|c|c|c}
    \hline
    $T/M$ & 700 & 800 & 900 & 1000 & 1100 & 1200
    \\
    \hline
    $r/10^3M$ & $1.4964$ & $1.7798$ & $1.9528$ & $2.1523$ & $2.3841$ & $2.6555$
    \\
    \hline
    $R/10^{-3}\cfrac{L^2}{M}$ & $0.6683$ & $0.5619$ & $0.5121$ & $0.4646$ & $0.4195$ & $0.3766$
    \\
    \hline
    \end{tabular}
    \caption{The position of source ${}_{-2}\hat{\mathcal{S}}^{[55]}$'s peak at different time slices corresponding to Fig. \ref{logS}.
    The coordinate speed of the peak is much less than the outgoing characteristic $c_+(R=0)=-1/8$.
    }
    \label{S_peak}
\end{table}

\section{Conclusion and outlook} \label{sec5}

In this work, we investigate second-order gravitational perturbations under the Schwarzschild spacetime that is perturbed by a transient electromagnetic wave packet.
Electromagnetically sourced nonlinear tails are found by numerically solving the inhomogeneous BPT equation.
Accurate numerical results are obtained by efficient numerical methods, including AnMR and time-symmetric integration. 
When $l\geq4$, the nonlinear tails of curvature perturbations decay as $ t^{-2l-2}$ at fixed spatial position, slower than the linear Price's law $t^{-2l-3}$, and $u^{-l-3}$ at null infinity, which are independent of mode coupling and initial data.
While $l=2$ and $3$, given that real matter is compact, the nonlinear tails decay at the same rate as the linear tails under the compact support initial data.
These tail behaviors are a bit different from the nonlinear tails sourced by first-order gravitational perturbations when $l=2$ and $3$, because the latter still decay as $ t^{-2l-2}$ \cite{Cardoso2024}.
This difference could be used to identify astrophysical origins of GWs in the multi-messenger observations, providing a novel and complementary mechanism to detect black holes in the Milky Way.

One direction for future efforts is to extend the current analysis to Kerr spacetime, where nonlinear tails may reveal richer phenomena, e.g., mode projections or intermediate “splitting” of decay rates \cite{Zenginoglu2014}.
Given the rise in computing costs, it is essential to benchmark the efficiency of various numerical methods.
In addition to the AnMR and time-symmetric integration methods used in this article, other approaches, such as the multi-domain spectral method, fully spectral method \cite{Macedo2014}, and spectral decomposition method \cite{Ansorg2016}, are also suitable for studying polynomial tails.

Masked by the stronger contribution of QNMs to ringdown waveform modeling, tails have generally been regarded as negligible for near-future GW detections. 
However, the tail effect could be significantly amplified in inspirals with high eccentricities, potentially detectable by upcoming detectors \cite{Albanesi2023,Carullo2023,DeAmicis2024a,Islam2025}.
The power law of these eccentricity-induced tails seems to exhibit competition between source-free tails and source-driven ones \cite{Islam2025}.
Thus, another direction for future efforts is to investigate the electromagnetically sourced nonlinear tails in such an extreme mass ratio inspiral system comprised of a neutron star and a supermassive BH.

\begin{acknowledgments}
This work is partly supported by the National Key Research and Development Program of China (Grant No. 2021YFC2203001).
This work is supported in part by the National Natural Science Foundation of China under Grants No. 12035016, No. 12075026, No. 12275350, No. 12375048, No. 12375058, No. 12361141825, No. 12447182 and No. 12575047.
\end{acknowledgments}

\appendix
\section{Tetrad and spin coefficients}\label{sec_tetrad}

The tetrad employed in this work reads
\begin{equation} \label{l}
    l^{\mu}= \frac{R^2}{L^4}
    \left\{
    4M^2,-\frac12(L^2-2MR),0,0
    \right\},
\end{equation}
\begin{equation}\label{n}
    n^{\mu}=
    \left\{
    2+\frac{4MR}{L^2},\frac{R^2}{L^2},0,0
    \right\},
\end{equation}
\begin{equation}\label{m}
    m^\mu = \frac{R}{\sqrt{2}L^2}
    \left\{
     0,0,-1,-\frac{\ii}{\sin\theta}
    \right\}.
\end{equation}
The unperturbed tetrad vectors (\ref{l}) and (\ref{n}) are chosen along the repeated principal null directions of the Weyl tensor, so that one can obtain
\begin{equation}
    \Psi^{(0)}_0 
    =
    \Psi^{(0)}_1
    =
    \Psi^{(0)}_3
    =
    \Psi^{(0)}_4
    = 0
\end{equation}
and 
\begin{equation}
    \kappa^{(0)}
    = 
    \sigma^{(0)}
    =
    \nu^{(0)}
    =
    \lambda^{(0)}
    =0 ,
\end{equation}
following from the Goldberg-Sachs theorem \cite{Goldberg2009}. 

The only non-zero Weyl scalar is 
\begin{equation}
    \Psi_{2}^{(0)}=-\frac{MR^{3}}{L^{6}},
\end{equation}
and the non-zero spin coefficients are
\begin{equation}
    \rho^{(0)} = - \frac{R(L^2-2MR)}{2L^4},
\end{equation}

\begin{equation}
    \mu^{(0)} = - \frac{R}{L^2},
\end{equation}

\begin{equation}
    \epsilon^{(0)} = \frac{M R^2}{2 L^4},
\end{equation}
and 
\begin{equation}
    \alpha^{(0)} =
    -\beta^{(0)} = 
    \frac{R \cot\theta}{2 \sqrt{2} L^2}.
\end{equation}
Finally, the vanishing spin coefficients are
\begin{equation}
    \tau^{(0)}=\varpi^{(0)}=\gamma^{(0)}=0.
\end{equation}

\bibliography{Tails.bib}
\end{document}